\documentclass[10pt]{article}

\usepackage{etex} 
\usepackage{pdfpages}
\usepackage{amsmath}
\usepackage{amssymb}
\usepackage{amsthm}
\usepackage{times}
\usepackage{subscript}
\usepackage{times}
\usepackage[latin1]{inputenc}
\usepackage[english]{babel}

\usepackage{hyperref}

\usepackage{colortbl}
\usepackage{txfonts}
\usepackage{xypic}
\usepackage{multirow}
\usepackage{textcomp}
\usepackage{calc}
\usepackage{stmaryrd}
\usepackage{marvosym}
\usepackage{pgf}
\usepackage{ifthen}
\usepackage{mdwlist}
\usepackage{enumerate}
\usepackage{paralist}
\usepackage[caption=false]{subfig}

\usepackage{tikz}
\usetikzlibrary{shapes.geometric}
\usetikzlibrary{arrows,automata}
\usetikzlibrary{decorations.pathmorphing}
\usetikzlibrary{decorations.markings}
\usetikzlibrary{decorations.shapes}
\usepackage{tikz-qtree}

\usepackage{algorithm}
\usepackage{algpseudocode}
\usepackage{calc}
\usepackage{linegoal}

\algnewcommand{\LineComment}[1]{\State \(\triangleright\) #1}

\algblockdefx[NAME]{Begin}{End} %
    {\textbf{begin}} %
    {\textbf{end}}

\newcommand*{\LongState}[1]{\State
\parbox[t]{\linegoal}{#1\strut}}

\DeclareMathAlphabet{\mathpzc}{OT1}{pzc}{m}{it}

\tikzset{
	path/.style={dotted},
	every edge/.style={draw,solid},
	normal/.style={solid},
}

\makeatletter
\def\th@plain{%
  \thm@notefont{}
  \itshape 
}
\def\th@definition{%
  \thm@notefont{}
  \normalfont 
}
\makeatother

\theoremstyle{definition}
\newtheorem{definition}{Definition}[section]
\newtheorem{example}{Example}[section]

\theoremstyle{remark}
\newtheorem{remark}{Remark}[section]

\newcommand{\der}{\leftarrow} 

\title{Approximate Query Answering in Inconsistent Databases (Minor Research Report)}
\author{Federica Panella\medskip\\
{\small DEIB - Politecnico di Milano}\\
{\small federica.panella@polimi.it}
}

\date{}

\begin{document}

\maketitle

\begin{abstract}
Classical algorithms for query optimization presuppose the absence of inconsistencies or uncertainties in the database and exploit only valid semantic knowledge provided, e.g., by integrity constraints. Data inconsistency or uncertainty, however, is a widespread critical issue in ordinary databases: total integrity is often, in fact, an unrealistic assumption and violations to integrity constraints may be introduced in several ways. 

In this report we present an approach for semantic query optimization that, differently from the traditional ones, relies on not necessarily valid semantic knowledge, e.g., provided by violated or soft integrity constraints, or induced by applying data mining techniques. Query optimization that leverages invalid semantic knowledge cannot guarantee the semantic equivalence between the original user's query and its rewriting: thus a query optimized by our approach yields approximate answers that can be provided to the users whenever fast but possibly partial responses are required. Also, we evaluate the impact of use of invalid semantic knowledge in the rewriting of a query by computing a measure of the quality of the answer returned to the user, and we rely on the recent theory of Belief Logic Programming to deal with the presence of possible correlation in the semantic knowledge used in the rewriting.
\end{abstract}

\section{Introduction}
\label{sec:introduction}

Data inconsistency or uncertainty is a widespread critical issue in ordinary databases. 
The integrity of a database -- in theory -- is ensured by declaring assertions (integrity constraints) that are required to hold in every instance of the database. Total integrity is often, however, an unrealistic assumption, and violations to the constraints may be introduced in several ways.
 
Inconsistencies may arise when multiple databases, with possibly different sets of integrity constraints, are merged; discrepancies may also occur in data warehouses or in replicated systems,
or whenever integrity checking is temporarily loosened, e.g., for uploading a backup.
Whenever integrity violations enter into the database, consistency is rarely restored again, as users and administrators usually deem it too expensive or even impractical to remove existing violations in compromised data.

Furthermore, the requirements specified for the consistency of a database may be modeled by hard integrity constraints, whose satisfaction must be invariably enforced, or soft integrity constraints, which are not strictly required to be valid, and thus may tolerate the presence of violations in  a database instance.

Semantic knowledge about a database instance might also be induced by applying data mining techniques, searching for frequent relationships among elements (attributes or predicates) of the schema: the discovered relationships define assertions on the examined database instance (e.g., in the form of association rules~\cite{Agrawal1993}) that may be interpreted as integrity constraints whose satisfaction is not fully complied with by the whole set of data.

In this report we focus on the problem of efficient approximate query answering in the presence of semantic knowledge about the database provided by not necessarily valid integrity constraints. This information is exploited to answer user's queries when fast and approximate answers are required. 

Our approach is based on the classical technique of Semantic Query Optimization (SQO) \cite{GrantMinker1992}, where semantic knowledge about the database is used to rewrite the user's queries into a semantically equivalent form (i.e., with the same set of answers) where redundant table scans and joins are possibly eliminated, so that the queries can be processed more efficiently. 
Differently from the classical approach, we allow also inconsistent semantic knowledge to be leveraged on to optimize the queries and reduce their processing time. The rewriting based on potentially violated integrity constraints, however, does not necessarily preserve the semantics of the original query; nevertheless, we provide bound guarantees on the degree of certainty of the answers that are retrieved in the presence of impaired semantic knowledge.

To this end, we assume that integrity constraints are annotated with a measure of their validity or their satisfaction in the current database instance, and we compute a bound on the approximation error induced by their use for the transformation of the original query.
Computing  the certainty degree of the answers raises, however, various issues, since the integrity assertions can be in general not independent and their measures of quality cannot be straightforwardly combined to yield an estimate of the impact of their use to optimize a query. 

We address these issues by relying on the recent theory of Belief Logic Programming (BLP)~\cite{WanKifer2009}, a model of quantitative reasoning that was specifically designed for dealing with correlation of elements of knowledge that can be inconsistent and not necessarily independent.
In BLP the contents of a database (consisting of facts and rules) is explicitly annotated with estimates of its degree of certainty, which represent the belief in its validity supported by possibly conflicting or inaccurate information sources.  
The likelihood of the elements of knowledge that can be inferred by submitting queries to the database is computed resorting to belief combination functions,  similarly to Dempster-Shafer's theory of evidence~\cite{Dempster1967,Shafer1976}.
In fact, belief combination functions are able to aggregate the annotations of the degree of certainty of various elements of knowledge, providing an estimate of the validity of the overall conclusions that can be derived from them, and properly heed the presence -- if any -- of correlation in the available data.

The contribution of this report is the extension of the BLP theory with the presence of annotated integrity constraints that we exploit for the semantic optimization of users' queries, yielding approximate answers that are easier to compute.
The impact of using uncertain semantic knowledge in the process of query answering is evaluated by  computing two measures of accurateness of the answers retrieved by the procedure: their correctness (i.e., the likelihood by which an answer to a query optimized by the use of integrity constraints is indeed an answer to the original query) and  we estimate the loss of completeness of the procedure (i.e., the likelihood by which an answer to the original query is not actually retrieved by the approximate query  answering algorithm). These measures are evaluated in two stages, to guide the procedure of semantic query optimization in the choice of a suitable rewriting of a query, and to provide the user with a characterization of the quality of the approximate answers returned to him after executing the rewritten query.
Since an exact evaluation of the accurateness of the procedure would actually undermine the benefits provided by query optimization, 
we compute a lower bound on the correctness of the procedure and  we estimate an upper bound on its lack of completeness.
To compute these bounds, we rely on belief combination functions of BLP theory to properly integrate the dependencies in the semantic knowledge exploited to optimize a query.

The outline of the report is as follows. Section~\ref{sec:preliminaries} introduces the basic notation used in this work and some preliminary background on the classical technique of SQO and on BLP theory. Section~\ref{sec:approximateQA} illustrates our approach for approximate query answering for a database consisting only of facts, while in the following section, Section~\ref{sec:approximateQAExtensions}, we extend the procedure to deal with general databases that include also rules. Section~\ref{sec:future-work} describes some extensions for future work and presents the campaign of experiments we plan to perform to validate our approach. Section~\ref{sec:relatedWork} presents some related works. 
Finally, Section~\ref{sec:conclusions} concludes the report.

\section{Preliminaries}
\label{sec:preliminaries}

In the following we introduce the notation and the basic logic definitions used in this work. We refer the reader to any classical text in logic programming and databases such as~\cite{CeriGottlobTanca1990,Nilsson1990} for additional background.

We consider a first-order language with infinite sets of symbols for \emph{predicates}, \emph{constants} and \emph{variables}; as usual for deductive databases, the language does not contain function symbols. The alphabet of the language consists also of logical connectives, quantifiers, parentheses and comma.
As a notational convention, symbols $p, q,\ldots$ denote predicates, $a, b,\ldots$  denote constants and $x, y,\ldots$ denote variables.

A \emph{term} is either a variable or a constant; terms are typically denoted by symbols $s, t$ and sequences of terms are represented as vectors, e.g., $\vec{t}$.

Each predicate has an associated nonnegative arity and the notation $p/n$ indicates that predicate $p$ has arity $n$.
Predicates applied to terms generate \emph{atoms} or \emph{atomic formulae}, that is, if $p$ is a predicate of arity $n$ and $t_1,\ldots, t_n$ are terms, then $p(t_1,\ldots, t_n)$ is an atom. Atoms are conventionally denoted by symbols  
$A, B,\ldots$

The set of \emph{(well-formed) formulae} is defined in the usual way. 

An occurrence of a variable $x$ is said to be \emph{bound} in a formula $F$ if either it is
the occurrence of $x$ in a quantifier $\forall x$ or $\exists x$ in $F$ or it lies within the scope of a
quantifier $\forall x$ or $\exists x$ in $F$. Otherwise, the occurrence is said to be \emph{free} in $F$.

A formula is $closed$ if it has no free variables.

A substitution is a finite set of pairs of terms $\{ x_1/t_1,\ldots,x_n/t_n \}$ (also denoted $\{\vec{x}/\vec{t}\}$) where each $x_i$ is a variable and each $t_i$ is a term and the same variable cannot be mapped to different terms. 
Let $dom(\sigma)$ denote the set $\vec{x}$ of variables in the substitution $\sigma = \{\vec{x}/\vec{t}\}$.
The application of a substitution $\sigma$ to a term (resp. formula) $E$ is denoted by $E \sigma$ and represents the term (resp. formula) constructed from $E$ where 
each free occurrence of a variable in $\vec{x}$ is simultaneously replaced by the corresponding term in $\vec{t}$.
We extend this notation also to sequences  of variables $\vec{y}$.

A formula or term is called \emph{ground} if it contains no variables.

A \emph{literal} is either an atom (positive literal) or the negation of an atom (negative literal).

Predicates can be distinguished into three distinct categories: \emph{intensional}, \emph{extensional}, and \emph{built-in} predicates.
The set of built-in predicates is $OP = \{\doteq, \neq, <, >, \leq,$ $\geq\}$.
Intensional and extensional predicates are collectively called \emph{database predicates}; atoms and literals are classified accordingly on the basis of their predicate symbol.

A \emph{clause} is a disjunction of literals. It can be written as a formula $A \der L_1 \land \ldots \land L_n$ where $A$ is a possibly missing atom and $L_1,\ldots, L_n$ ($n \geq 0$) are literals; as usual, all the variables are implicitly universally quantified. 
$A$ represents the \emph{head} of the clause and $L_1 \land \ldots \land L_n$ forms the \emph{body} of the clause.
An empty head can be seen as a nullary connective $false$; symmetrically an empty body can be understood as a nullary connective $true$.

A \emph{Horn} clause is a disjunction of literals of which at most one is positive: it can be written as a formula  $A \der A_1 \land \ldots \land A_n$ where $A_1,\ldots, A_n$ are atoms and the head can be possibly empty. 
A \emph{definite} clause is a Horn clause with a non empty head.
A \emph{unit} clause is a definite clause with an empty body. 
A \emph{fact} is a ground unit clause whose head is extensional. 
A \emph{rule} is a clause whose head is intensional.

A clause is said to be \emph{range-restricted} if every variable appears in a positive database literal in the body.

\begin{definition}[Deductive database]
A \emph{deductive database} consists of three components: a finite set of facts, the \emph{extensional database} (EDB); a finite set of range-restricted rules,  the \emph{intensional database} (IDB); a finite set of clauses called \emph{integrity constraints}, the \emph{constraint theory} (IC).
\end{definition}

\noindent In this report we consider integrity constraints expressed as range-restricted clauses.

To define the semantics of the first-order language for a deductive database, we introduce the classical definitions of Herbrand base, interpretation and model.

\begin{definition}[Herbrand universe, Herbrand base]
Let $D$ be a database. The \emph{Herbrand universe} $U_D$ of $D$ is the set of all constants occurring in $D$.
The \emph{Herbrand base} $H_D$ of $D$ is the set of all ground atomic formulae over $U_D$.
\end{definition}

\begin{definition}[Herbrand interpretation]
A \emph{Herbrand interpretation} of a database $D$ is any subset $I$ of $H_D$.
\end{definition}

\begin{definition}[Semantics of (closed) formulae]
Let $I$ be a Herbrand  interpretation for a database $D$. A closed formula $F$ is \emph{true} (respectively \emph{false}) in $I$, written $\models_I F$ (respectively  $\not\models_I F$), according to the following definition.
For any $I$, $\models_I true$ and $\not\models_I false$. For a ground atom $A$, $\models_I A$ iff $A \in I$.  
For any closed formulae $F$, $F_1$, $F_2$, $\models_I \neg F$ iff $\not\models_I F$; $\models_I F_1 \land F_2$ iff $\models_I F_1$ and $\models_I F_2$; $\models_I F_1 \vee F_2$ iff $\models_I F_1$ or $\models_I F_2$;  $\models_I F_1 \leftarrow F_2$ iff $\models_I F_1$ or $\not\models F_2$;  $\models_I \forall x F$ (respectively, $\models_I \exists x F$) iff for all (respectively, some) constant $c$ in $D$, $\models_I F\{x/c\}$.
\end{definition}

\begin{definition}[Validity]
A formula is \emph{valid} iff it is true in every interpretation.
\end{definition}

We assume the validity of a basic set of formulae that define the semantics of the built-in predicates.

The following \emph{free equality axioms} assure that the predicate of equality $\doteq$ is interpreted as the identity relation in every Herbrand interpretation. All the variables in these formulae are universally quantified, and $p$ denotes any predicate of $D$.
\begin{align*}
& x \doteq x  & (\text{reflexivity})  \\
& y \doteq x \leftarrow  x \doteq y  & (\text{simmetry})\\
& x \doteq z \leftarrow x \doteq y \land y \doteq z  & (\text{transitivity})\\
& p(y_1,\ldots, y_n) \leftarrow p(x_1,\ldots, x_n) \land x_1 \doteq y_1 \land \ldots x_n \doteq y_n & (\text{substitutivity of }\doteq)
\end{align*}
The following set of formulae enforces built-in predicate $<$ to be a partial order.
\begin{align*}
& \neg (x < x)  & (\text{irreflexivity})  \\
& x < z \leftarrow x < y \land y < z  & (\text{transitivity})
\end{align*}
The other predicates in $OP$ can be clearly expressed in terms of the previous twos: for every two terms $t_1, t_2$, $t_1 \neq t_2$ is an abbreviation for $\neg (t_1 \doteq t_2)$; $t_1 \leq t_2$ (respectively, $t_1 \geq t_2$) is a shorthand notation for $t_1 < t_2 \vee t_1 \doteq t_2$ (respectively, $t_1 > t_2 \vee t_1 \doteq t_2$), whereas $t_1 > t_2$ corresponds to $\neg (t_1 \leq t_2)$.

A common assumption for deductive databases is that they include a further set of formulae, namely the  \emph{unique name axioms}, which state that two constants cannot denote the same element in $U_D$:
\[
\neg(c_1 \doteq c_2) \land \neg(c_1 \doteq c_3) \land \ldots \land \neg(c_{n-1} \doteq c_n)
\]
where $c_1,c_2,\ldots,c_n$ are all the constants occurring in the database.


\begin{definition}[Herbrand model]
An interpretation $I$ of a database $D = \langle IDB, EDB,$ $ IC \rangle$  is a \emph{(Herbrand) model} of $D$ if $\models_I C$ for every clause $C \in EDB \cup IDB$.
\end{definition}

In the following we will deal with a standard restricted class of databases, namely hierarchical databases, defined next.

\begin{definition}[Dependence relation]
For a database $D$ and each  pair of predicates $p$ and $q$, $p$ is said to \emph{depend} on $q$ if 
either there is a clause $H \der B$ in $D$ such that $p$ occurs in $H$ and $q$ in $B$, or
there is a predicate $r$ such that $p$ depends on $r$ and $r$ depends on $q$.
\end{definition}

\begin{definition}[Hierarchical database]
A database $D$ is \emph{hierarchical} if no predicate depends on itself.
\end{definition}

Among the possibly several (minimal) models of a database, we will consider as \emph{canonical model} its \emph{well-founded model}, which coincides with the \emph{standard model} for stratified databases (for further details on the definition of the well-founded or the standard model see, e.g.,~\cite{Nilsson1990}).

For a closed formula (or a set of closed formulae) $F$ and a database $D$, we will denote by $D \models F$ the truth of $F$ in the standard model $\mathcal{M}_D$ of $D$.

Also, we follow the convention of referring to the standard model to define the consistency of a database.

\begin{definition}[Consistency]
A database $D$ is \emph{consistent} iff $D \models IC$.
\end{definition}

\begin{definition}[Query]
A \emph{query} (or \emph{goal}) on a database $D$ is a clause with an empty head.
\end{definition}

In the sequel of this report we will represent a query as a new intensional predicate of the database, introducing a corresponding rule having as body the body of the clause, i.e., for a query $\der Q(\vec{x})$ on a database $D$ we will define a new predicate $q/n$ and a range-restricted rule  in the $IDB$ $q(\vec{y}) \der Q(\vec{x})$, where $\vec{y}$ is the $n$-tuple of \emph{distinguished} (or \emph{output}) \emph{variables} of $q$.

\begin{definition}[Extension of a predicate]
The \emph{extension} of a database predicate $p/n$ in a database $D$ is defined as the set of $n$-tuples $E^{p}_{D} = \{\vec{a} : \ D \models p(\vec{a}) \}$. The extension of a (well-formed) formula is defined analogously.
\end{definition}

\begin{definition}[Answer to a query]
An \emph{answer} to a query $q$ on a database $D$ is a tuple in the extension $E^{q}_{D}$ of the intensional predicate $q$.
\end{definition}

Two queries are \emph{(semantically) equivalent} if they have the same answers on all databases.

\subsection{Semantic Query Optimization}
\label{sec:SQO}

In this section we briefly recall the classical approach to Semantic Query Optimization, presented, e.g., in~\cite{GrantMinker1990}.

As previously mentioned, the aim of SQO is to exploit integrity constraints to optimize a query on a consistent database into a (semantically) equivalent and more efficient form.
The optimization consists of two stages: a \emph{semantic compilation} phase and a \emph{semantic transformation} phase.

The first stage is based on the technique of \emph{partial subsumption},  which produces fragments of the integrity constraints -- called \emph{residues} -- that are associated with the extensional and intensional predicates of the database and are used during the subsequent phase to simplify the query.
Partial subsumption is an extension of the well-known concept of subsumption, which defines a relationship of entailment between two clauses. Basically, 
a clause $C$ is said to \emph{subsume} a clause $D$ if there exists a substitution $\sigma$ such that every literal in $C \sigma$ is also in $D$.
Partial subsumption consists in the application of the standard SLDNF-based algorithm for checking subsumption where we consider as clause $C$ an integrity constraint and as clause $D$  the body of a rule. If the integrity constraint subsumes the body of the rule, the algorithm yields a refutation tree ending with a null clause, meaning that the rule violates the constraint. This should be an unexpected result, as in general we refer to databases with non-conflicting components. Thus, in general  the subsumption algorithm will yield a refutation tree terminated by a non-null clause: this clause (called residue) represents a condition that must be true -- and thus can be asserted in the database -- when the body of the rule and the integrity constraint are satisfied.

The phase of semantic compilation first processes every integrity constraint to avoid the presence of constants in its atoms: each constant occurring in an atom is replaced with a new variable, and an equality restriction of the constant and the new variable is introduced in the body of the constraint.
Then it applies partial subsumption to each integrity constraint and either the body of a rule or the body of dummy rules $r \der r$ or $\neg r \der \neg r$   where $r$ is an extensional predicate. The resulting residues are associated with the corresponding predicate (respectively, the intensional predicate defined by the rule or the positive or negated extensional predicate of the dummy rules).

To illustrate the classical procedure, we consider an example of database involving papers, authors and conferences, which will be used as running example throughout this report.
\begin{example}
\label{ex:dbBook}
A bibliography deductive database (whose predicates are partly based on an example from~\cite{Aftrati2012}) has the predicates:\\

\noindent Paper(oid, subject)\\
Author(surname, age, nationality)\\
AuthorPaper(surname, oid)\\
Authoritative(surname, subject)\\
Conference(name, subject, venue)\\
PCMember(surname, conference)\\
Bestseller(oid)\\
Award(conference, oid)\\

\noindent that are all extensional. In particular, the predicate $PCMember(surname, conference)$ denotes a program committee member of a conference, whereas $Authoritative(surname,$ $ subject)$ denotes an authoritative author on a given subject. The extensional database consists of corresponding ground atoms, not shown here.
 
The database has an integrity constraint, stating that a paper cannot have an $oid$ with value $0$.

\[IC_0 :  \der Paper(0, subject)\]

The integrity constraint is first preprocessed, introducing a new variable, resulting in the following clause:

\[IC_0' :  \der Paper(x, subject), x=0\]

Then, for the dummy rule $Paper(oid, subject) \der Paper(oid, subject)$, the subsumption algorithm between $IC_0'$ and $D=  \der Paper(oid, subject)$ consists of the following steps: the body of the rule is grounded by a substitution $\theta$ that replaces each variable with a new constant; here $\theta = \{oid/k_1, subject/k_2\}$. The clause $D \theta =  \der Paper(oid, subject) \theta$ is negated, obtaining $\neg D \theta = Paper(k_1, k_2) \der$, and the  algorithm  yields the fragment $\der k_1 = 0$. The grounding substitution is reverted to generate the residue $\der oid = 0$ for the predicate $Paper(oid, subject)$.\qed
\end{example}

Note that, in the classical approach, the integrity constraints are assumed to be expressed only by extensional predicates: if this is not the case, the integrity constraints are initially preprocessed so that the intensional predicates occurring in their definition are elaborated into solely extensional ones.
The expansion cannot be performed in the case of negated intensional predicates whose unfolding introduces a negated existentially quantified variable (or, otherwise, existential quantifiers should be explicitly indicated in the constraint).

The second phase, semantic transformation, optimizes a query by exploiting the residues associated with the predicates occurring in its body.

First, a query is processed expanding each possible intensional predicate in its body, obtaining a set of queries consisting of solely extensional predicates and such that the generated set is semantically equivalent to the original query.
The residues of these predicates can be exploited to rewrite each query into an equivalent simplified form.
In fact, if the body of a residue $H \der B$ subsumes a query with a substitution $\sigma$, its head $H \sigma$ can be asserted in the database and may be used to transform the query in one of the following ways: either by removing, if present, the literal $H \sigma$ (hence, it allows the elimination of a join) or by introducing $H \sigma$ into the body of the query (the introduction of the literal can be useful if it has a small extension and is joined with the conjunction of predicates with large extensions already present in the query) or, if the head is a built-in atom, introducing a selection criterion (restriction) for a variable, which may be convenient whenever there is a query predicate sorted on this variable and the processing of the query may thus be limited to the values of the variable in the selected range.
Also, if the head of a residue that subsumes a query is empty, there is no need to perform the query at all, as its set of answers must be empty.

\subsection{Belief Logic Programming}
\label{sec:BLP}

This section introduces the basic concepts of the theory of Belief Logic Programming that will be used in the sequel of this report. For more details see~\cite{WanKifer2009}.

BLP is a form of quantitative reasoning able to deal with the presence of uncertain elements of knowledge derived from  non-independent and, possibly, contradictory information sources. The uncertainty in knowledge representation is modeled by degrees of beliefs, in a similar way as in Dempster-Shafer theory of evidence.

A database is represented as a set of \emph{annotated rules}, called \emph{belief logic program} (\emph{blp}). Each annotated rule is denoted by:
\[
\begin{array}{ccc}
[v,w] & H \der Body\\
\end{array}
\]
where $H$ is a positive atom and $Body$ is a Boolean combination of atoms, i.e., a formula composed out of atoms by conjunction, disjunction, and negation.
The annotation $[v,w]$ is called a \emph{belief factor}, where $v$ and $w$ are real numbers with $0 \leq v \leq w \leq 1$.
The intuitive meaning of this rule is that if $Body$ is true, then this rule supports the truth of $H$ to the degree $v$ and $\neg H$ to the degree $1-w$. The difference $w - v$ represents the information gap (or the degree of ignorance) with regard to $H$.

An \emph{annotated fact} $H$ is represented by an annotated rule of the form 
$[v,w] \ H \der true$, written simply as $[v,w] \ H$.

In the following we will refer simply to rules and facts instead of annotated rules and annotated facts, with the usual convention that the predicate in the head of a rule is intensional and for a fact is extensional.

The dependence relation between atoms in a blp can be defined analogously to the definition presented before for the predicates of a classical database.

BLP uses explicit negation (or strong negation) rather than negation as failure. Thus, if the truth of an atom $A$ cannot be deduced from the rules and facts of the blp, we cannot infer that the negation of $A$ holds, but only that there is no evidence that $A$ is true.

To specify the semantics of a blp, we first introduce the concept of belief combination functions; then, we define the notions of interpretation and model.

\begin{definition}
Let $N$ be the set of all sub-intervals of $[0,1]$. A function $\phi: N \times N \rightarrow N$ is a \emph{belief combination function} if it is associative and commutative.
\end{definition}

Due to the associativity of the function $\phi$, we can extend it from two to three or more arguments as follows: $\phi([v_1,w_1], \ldots, [v_k,w_k]) = \phi( \phi([v_1,w_1], \ldots, [v_{k-1}, w_{k-1}]),[v_k,w_k])$. Furthermore, the order of arguments in a belief combination function is immaterial, since such functions are commutative. Thus, we can write them as functions on a multiset of intervals, e.g., $\phi(\{[v_1,w_1],...,[v_k,w_k]\})$.

As a short notation, we denote by $\phi_V$ and $\phi_W$ the projection of function $\phi$ on the left and right value, respectively, of the interval it generates, i.e., if $R = \{[v_1,w_1],...,[v_k,w_k]\}$ and $ \phi(R) = [v_{k+1}, w_{k+1}]$, then $\phi_V(R) = v_{k+1}$ and $\phi_W(R) = w_{k+1}$.

For convenience, we also extend $\phi$ to the nullary case and the case of a single argument as follows: $\phi() = [0, 1]$ and $\phi([v, w]) = [v, w]$.

There are many belief combination functions, which can be properly used for different application domains and different types of data. Three of the most commonly used are:
\begin{itemize}
\item Dempster's combination rule: 
\begin{itemize}
\item $\phi^{DS}([0,0],[1,1]) = [1,1]$;
\item $\phi^{DS}([v_1,w_1], [v_2,w_2]) = [v,w]$ if $\{[v_1,w_1], [v_2,w_2]\} \neq \{[0,0],[1,1]\}$, where $v = \frac{v_1 w_2 + v_2 w_1 - v_1 v_2}{K}, w = \frac{w_1 w_2}{K}$, and $K = 1 + v_1 w_2 + v_2 w_1 - v_1 - v_2$ ($K \neq 0$)
\end{itemize}
\item Maximum: $\phi^{MAX}([v_1,w_1],[v_2,w_2])=[max(v_1,v_2),max(w_1,w_2)]$.
\item Minimum: $\phi^{MIN}([v_1,w_1],[v_2,w_2])=[min(v_1,v_2),min(w_1,w_2)]$.
\end{itemize}

Given a blp $P$, the definitions of Herbrand Universe $U_P$ and Herbrand Base $H_P$ of $P$ are the same as in the classical case. 
We refer, however, to a three-value logic and we define a \emph{truth valuation} over a set of atoms $\alpha$ as a mapping from $\alpha$ to $\{t,f,u\}$. The set of all possible valuations over $\alpha$ is denoted by $TVal(\alpha)$.

A \emph{truth valuation} for a blp $P$ is a truth valuation over $H_P$. Let $TVal(P)$ denote the set of all the truth valuations for $P$, so $TVal(P) = TVal(H_P)$.

If $\alpha$ is a set of atoms, we will use $Bool(\alpha)$ to denote the set of all Boolean formulas constructed out of these atoms (i.e., using conjunction, disjunction and negation).

\begin{definition}
Given a truth valuation $I$ over a set of atoms $\alpha$ and a formula $F\in Bool(\alpha)$, $I(F)$ is defined as in Lukasiewicz's three-valued logic: $I(A \vee B)$ $= max(I(A),$ $ I(B))$, $I(A \land B) = min(I(A), I(B))$, and $I(\neg A) = \neg I(A)$, where $f < u < t$ and $\neg t = f$, $\neg f = t$, $\neg u = u$. We say that $\models_I F$ if $I(F) = t$.
\end{definition}

\begin{definition}
A \emph{support function} for a set of atoms $\alpha$ is a mapping $m_{\alpha}$ from $TVal(\alpha)$ to $[0, 1]$ such that 
$\sum_{I \in TVal(\alpha)} m_{\alpha}(I) = 1$.
The set of atoms $\alpha$ is called the \emph{base} of $m_{\alpha}$. 
A \emph{support function} for a blp $P$ is a mapping $m$ from $TVal(P)$ to $[0, 1]$ such that
$\sum_{I \in TVal(P)} m(I) = 1$.
\end{definition}

\begin{definition}
A mapping $bel : Bool(H_P) \rightarrow [0, 1]$ is called a \emph{belief function} for $P$ if there exists a support function $m$ for $P$, so that for all $F \in Bool(H_P)$\\
$bel(F) = \sum_{I \in TVal(P) \ : \ \models_I F} m(I)$.
\end{definition}

Belief functions can be understood as interpretations of belief logic programs. 

\noindent We now introduce the notion of model.

Let $P$ be a blp, $I$ a truth valuation  and $X$ an atom  in $H_P$, and let $P(X)$ be the set of rules or facts in $P$ having $X$ as head. We define the \emph{P-support for $X$ in $I$}, $s_P(I,X)$, which measures the degree by which $I$ is supported by the rules or facts in $P(X)$: let $[v,w]$ be the result of the application of the combination function $\phi$ to the set of rules or facts in $P(X)$ whose body is true in $I$ (recall that for an empty set, $\phi()=[0,1]$).
If $X$ is true in $I$, then $s_P(I,X) = v$ (i.e., it represents the combined belief in $X$).
If $X$ is false in $I$, then $s_P(I,X) = 1-w$ (i.e., it represents the combined disbelief in $X$).
Otherwise, $s_P(I,X) = w - v$.

The support to $I$ from all the atoms in $H_P$ is defined by the function
$\hat{m}_p(I) = \prod_{X \in H_P} s_P(I,X)$,
which is also a support function.

Finally, we define: 
\begin{definition}
The \emph{model} of a blp $P$ is
\begin{equation}\label{eq:modelBLP}
model(F) = \sum_{I \in TVal(P) \ : \ \models_I F} \hat{m}_p(I), \quad \text{ for any } F \in Bool(H_P).
\end{equation}
\end{definition}

The model of $P$ is a belief function that assigns to each formula $F \in Bool(H_P$) the degree of certainty of $F$ in the blp, and it is unique for each blp.

\begin{remark}
This semantics of BLP is non-monotonic.
For instance, consider the combination function $\phi^{DS}$ and two facts $r_1: [0.4, 0.4] \ X$ and $r_2: [0.8, 0.8] \ X$ for the atom $X$. Consider also the programs $P_1 = \{ r_1\}, P_2 = \{ r_2 \}, P_3 = \{r_1, r_2\}$, and let $bel_i$ be the model of $P_i$ ($1\leq i \leq 3$). Since $\phi^{DS}([0.4, 0.4],[0.8, 0.8]) = [v,w]$ with $bel_3(X)=v < 0.8$, then adding $r_1$ to $P_2$ reduces the support of $X$.
\end{remark}

\subsubsection{Query answering for BLP}

We  briefly describe the query answering algorithm for BLP presented in~\cite{WanKifer2009}.

The algorithm provides an effective procedure to compute an answer to a query on a blp $P$, along with the degree of certainty (given by the model of the blp) by which it is supported by $P$. 
The algorithm is based on a fixpoint semantics for the notion of model, equivalent to the declarative one  
we presented before, and that we introduce in the following of this section along with some preliminary definitions.

\begin{definition}
Given a truth valuation $I$ over a set of atoms $\beta$, the \emph{restriction} of $I$ to $\alpha \subseteq \beta$, denoted $I|_{\alpha}$, is the truth valuation over $\alpha$ such that, for any $X \in \alpha$, $I|_{\alpha}(X) = I(X)$.
\end{definition}

\begin{definition}
Let $I_1$ be a truth valuation over a set of atoms $\alpha$ and $I_2$ be a truth valuation over a set of atoms $\beta$  disjoint from $\alpha$. The \emph{union} of $I_1$ and $I_2$, denoted $I_1 \uplus I_2$, is the truth valuation over 
$\alpha \cup \beta$ such that $(I_1 \uplus I_2)|_{\alpha} = I_1$ and $(I_1 \uplus I_2)|_{\beta} = I_2$.
\end{definition}

For a blp $P$ we denote by $SF(P)$ the set of all support functions for $P$: $SF(P) = \{m \mid  m$ is a support function for $\alpha \in H_P\}$. 
We define an operator $\hat{T}_P: SF(P) \rightarrow SF(P)$ for $P$ as follows.

\begin{definition}
Let $P$ be a blp.
For any support function $m$ in $SF(P)$, let $\alpha$ be the base of $m$ and let
\[
dep(\alpha) = \{ X \mid X \in H_P \setminus \alpha, \text{ and every atom $X$ depends on is in } \alpha \}.
\]
$\hat{T}_P(m)$ is a support function over the set of atoms $\alpha \cup dep(\alpha)$ such that
for every $I_1 \in TVal(\alpha)$ and $I_2 \in TVal(dep(\alpha))$,
\begin{equation}\label{eq:operatorTp}
\hat{T}_P(m)(I_1 \uplus I_2) = m(I_1) \cdot \prod_{X \in dep(\alpha)} Val(\phi_{X}(BF(X,P,I_1)),I_2(X))
\end{equation}

\noindent where $BF(X,P,I)$ is the multiset of the belief factors of all the rules or facts in $P$,
which have atom $X$ in head and whose bodies are true in $I$, and
$Val([v,w],\tau) = \left\{  
\begin{array}{ll}
v  & \text{if } \tau = t \\
1-w  & \text{if } \tau = f \\
w-v  & \text{if } \tau = u \\
\end{array} \right.$ 
\end{definition}

Basically, the operator $\hat{T}_P$ extends the definition of a support function $m$ with base $\alpha$ to a larger base, $\alpha \cup dep(\alpha)$, closed under the dependence relation between atoms. The extended support function $\hat{T}_P(m)$ assigns to each truth valuation $I_1 \uplus I_2$  a value of support given by a fraction of the support $m(I_1)$ that takes into account the contribution to the degree of certainty yielded by the new component added to the base. 
Each such component is assigned a fraction specified by $\prod_{X \in dep(\alpha)}$. We consider the product because under a fixed truth valuation $I_1$ over $\alpha$, the set of rules and facts that fire in $I_1$ and have different heads
are treated as independent and each member of the product represents the
contribution of each particular head $X \in dep(\alpha)$. In the contribution of a
particular head, $X$, $\phi_{X}(BF(X,P,I_1))$ is the combined belief factor $[v, w]$ of the rules or facts that support $X$ in $I_1$.
$Val$ yields the degree of belief in $X$ or the degree of belief in $\neg X$ or the uncertainty gap, depending on whether $X$ is true, false or uncertain in $I_2$.

The query answering algorithm for BLP  computes the model of a blp $P$  as a fixpoint of the operator $\hat{T}_P$ (which always exists). 

\begin{definition}
Let $P$ be a blp and $m_{\omega}$ be the fixpoint of $\hat{T}_P$. The model of $P$ is the following belief function:
\begin{equation}\label{eq:modelBLP-fixpoint}
model(F) = \sum_{I \in TVal(P) \ : \ \models_I F} m_{\omega}(I), \quad \text{ for any } F \in Bool(H_P).
\end{equation}
\end{definition}

and it is equivalent to the definition of model in Equation~\ref{eq:modelBLP}.

In the following we introduce first the query answering algorithm for ground blps and queries, then we present the generalization of the algorithm to non-ground blps and queries.

A ground query to a blp $P$ is a statement of the form $\der G$, where $G \in Bool(H_P)$; an answer to the query is the belief by $P$ in $G$.

For convenience, we add the query to the rules of the blp so that we can refer to it as an atom, i.e.,  given a blp $P$ and a ground query $\der G$, we denote by $P_g$ the set composed of all the rules and facts in $P$ plus the additional rule $[1, 1] \ g \der G$, where $g$ is a new ground atom. The belief in $G$ by $P$ is equivalent to the belief in $g$ by $P_g$.
We also assume that each rule and fact $R$ of $P$ is associated with a new predicate denoted $ID_R$.

The query answering algorithm is based on the use of some suitable data structures, namely dependency DAGs and proof DAGs, defined next.

\begin{definition}
The \emph{dependency DAG}, $H$, of a ground blp $P$ is a directed acyclic bipartite graph whose nodes are partitioned into a set of \emph{atom nodes} (a-nodes, for short) and \emph{rule nodes} (r-nodes, for short).
\begin{itemize}
\item For each atom $A$ in $P$, $H$ has an a-node labeled $A$.
\item For each rule or fact $R$ in $P$, $H$ has an r-node labeled $ID_R$.
\item For each rule or fact $R$ in $P$, an edge goes from the r-node labeled $ID_R$ to the a-node labeled with the atom in $R$'s head.
\item For each rule $R$ in $P$ and each atom $A$ that appears in $R$'s body, an edge goes from the a-node labeled $A$ to the r-node labeled $ID_R$.
\item Each edge that goes from an r-node, labeled $ID_R$, to an a-node is labeled with the belief factor associated with the rule or fact $R$.
\end{itemize}

\end{definition}

In dependency DAGs, if there is an edge from node $A$ to a node $B$, we call $A$ a child node of $B$ and $B$ a parent node of $A$.

The \emph{proof DAG} of $P$ for a ground query $g \der G$ is defined as  a subgraph of the dependency DAG for $P$ that contains all and only the nodes connected directly or indirectly to the a-node labeled $g$.

A \emph{pruned proof DAG} is obtained from a proof DAG by discarding redundant rule and atom nodes, i.e. it does not contain rules that are never triggered nor atoms that are supported by no rule or fact.

The query answering algorithm for ground queries computes the degree of belief in $g$ by building, incrementally, the belief function $model$ for the blp in Equation~\ref{eq:modelBLP-fixpoint}, starting from a belief function defined on an empty base $B$ and progressively expanding the base with nodes in $dep(B)$ until it includes the atom $g$.
The algorithm receives as input  a proof DAG of the blp for the query.
Initially, an empty base $B$ is extended to include a node without predecessors in the graph (i.e., for which the combined belief is already known) and the algorithm computes the model $m$ for this singleton base. Then the base is iteratively expanded to include a new node in a post-order traversal of the graph and the model function is updated to cover the new component, i.e., if $X$ is the new node added to the base $B$, the model is mapped to $m' = \hat{T}_P(m)|_{B \cup X}$, where for any $I_1 \in TVal(\alpha), I_2 \in TVal(\{X\})$,
$m'(I_1 \uplus I_2) = m(I_1) \cdot Val(\phi_{X}(BF(X,P,I_1)),I_2(X))$.
The algorithm ends when, after visiting the whole graph, the base includes also the node labeled $g$, and returns the value $model(g)$.
As a side remark, as regards some complexity issues of the algorithm, at every step the base is not monotonically expanded, but each time the base is enlarged, the nodes which are no longer needed for the rest of the computation are discarded.
The pseudo-code of the algorithm is presented in~\cite{WanKifer2009}.

We now consider non-ground programs and non-ground queries.

Let $P$ be a blp and $\der G$ be a non ground query, where $G$ is a Boolean combination of non-ground atoms.
By introducing in the blp a new rule for the query as before, we can assume that all non-ground queries are
 singleton non-ground atoms.
The query answering algorithm determines the ground tuples in the extension of $G$ in which the blp has a positive value of belief and returns to the user the set of these atoms together with the belief by the blp in them.

The algorithm is based on a procedure in the style of SLD-resolution. 
In general, SLD-resolution can only be applied to programs whose rules do not have disjunctions in the body, while blps might have such disjunctions. However, under the assumption that we consider combination functions $\phi$ such that $\phi([1,1],[1,1] = [1,1])$, every blp can be transformed into an equivalent blp without disjunctions in the body of the rules. Thus, in the following we will refer to rules expressed as clauses.

By applying a SLD-resolution procedure, the algorithm builds a proof-structure for the query, a SLD-tree, defined next.

\begin{definition}
An \emph{SLD-tree} for a blp $P$ and a query $\der G$ (both $P$ and $G$ can be non-ground) is built by applying the following steps:
\begin{itemize}
\item Add a node labeled with $\der G$ as root.
\item Repeat the following until no node can be added.
Let $o$ be a leaf node labeled with $\der G_1 \land \ldots \land G_n$, for every rule $R$ in $P$ of the form $[v,w] \ A \der B_1 \land \ldots \land B_m$ such that $G_1$ and $A$ or $G_1$ and $\neg A$ unify with the most general unifier (mgu) $\theta$,
\begin{itemize}
\item Add a node $o'$ labeled with $\der (B_1 \land \ldots \land B_m \land G_2 \land \ldots \land G_n)\theta$;
\item Add an edge from $o$ to $o'$ and label that edge with a triple $\theta, ID_R, Vars \theta$, where $ID_R$ is the identifier of $R$ and $Vars$ is the list of variables in
$R$.
\end{itemize}

In a special case when $o$ above is labeled with $\der  G_1$ (has only one atom) and $R$ above is a fact that unifies with $G_1$ or $\neg G_1$, $o'$ becomes a leaf node labeled with the empty clause. We call it a success node.

\item Delete the nodes and edges that are not connected to any success node.
\end{itemize}
\end{definition}

Then, the algorithm derives from the SLD-tree a set of pruned proof DAGs, one for each of the answers found by the procedure (i.e., for each ground tuples in the extension of $G$ in which the blp has a positive value of belief). Notice that, being the proof DAGs pruned, the algorithm allows us to avoid  building the whole dependency graph of the program to answer non-ground queries.
The answers to the query, along with the belief in them, are then computed by applying on these proof DAGs the previously described query answering algorithm for ground queries. 

\section{Approximate query answering}
\label{sec:approximateQA}

In this section we describe our approach for efficient query processing, which computes fast approximate answers to queries.

We rely on the classical technique of SQO, which exploits semantic knowledge about a database, i.e. integrity constraints, to improve performance of query answering.

In the classical scenario of SQO a query is transformed into an equivalent form that may be answered with a smaller processing cost. The basic assumption underlying the validity of the transformation is that the semantic properties of the data are always satisfied: in fact, the equivalence between the set of answers to the original query and to the rewritten one is not necessarily preserved if the integrity constraints are not invariably satisfied in the intended model of the database.

Still, we argue that uncertain semantic knowledge can be nonetheless effectively exploited for query optimization.
Unlike the classical approach to SQO, herein we refer to a scenario where the set of integrity constraints of a database does not hold with certainty and we show how to adapt the standard technique of SQO to deal with not necessarily valid semantic knowledge.
Although query rewriting based on uncertain integrity constraints cannot be semantics-preserving, the answers to the transformed query represent an approximation of the answers to the original one, which can be computed more efficiently, and we can evaluate the quality of the answers retrieved while processing the new query by providing a bound guarantee on the correctness and completeness of the optimized -- but not exact -- query answering procedure.

Semantic query optimization in the presence of uncertain semantic knowledge has to deal with a few issues, which we shall examine in the sequel of this section.

A first aspect concerns the characterization of the quality of the semantic knowledge about the database, provided by integrity constraints, in terms of its accurateness or ``certainty''.
Another aspect regards how to combine knowledge derived from integrity constraints that can be uncertain and possibly correlated, providing an evaluation of the impact of their use in the procedure of query optimization.
Yet another issue involves the analysis of the necessary trade-offs in the definition of a rewriting for a query submitted by a user, which must take into account both the benefits deriving from the uncertain knowledge to obtain a query that is easier to process and the downsides of reducing the accurateness of the approximate answer retrieved by executing the transformed query.

Thus, in this section we will
\begin{itemize}
\item provide a definition of the ``degree of certainty'' of the integrity constraints
\item adapt the classical approach of SQO to deal with the presence of uncertain semantic knowledge. In particular, we will associate annotations of certainty with the residues derived from the integrity constraints, and we will exploit them to optimize a query.
Since a simplified query will yield, in general, approximate answers, we describe two measures of correctness and completeness to evaluate the quality of a rewriting (which depends on the type of the rewriting, i.e. if it removes or inserts atoms, and on the certainty of the residues used to perform the transformation). We conclude by showing a simple procedure to compute these estimates to select a proper rewriting.
\end{itemize}

\subsection{Degree of certainty of semantic knowledge}
\label{sec:SQO-degree}

As regards the first issue, we consider two possible interpretations for the quality or degree of certainty of an integrity constraint.

Let $D$ be a database and $H_D$ its Herbrand base.
Let $C$ be the integrity constraint $A \der L_1, \ldots, L_n $ ($n\geq 1$).

The degree of certainty of $C$ can be interpreted as the likelihood that $C$ is satisfied in a database model of $D$ selected uniformly at random on $H_D$.
If the set of constants is finite, we can define the degree of certainty of $C$ as follows.

\begin{definition}
\label{def:degree-certainty-D}
The \emph{degree of certainty} of $C$ in $D$ is defined as the fraction of Herbrand models $I$ on $H_D$ where if 
$\models_I \land_{i=1}^{n} L_i$ then $\models_I A$.
\end{definition}

The degree of certainty of an integrity constraint can also be defined with reference to a specific model of a database, e.g. the standard model.

\begin{definition}
\label{def:degree-certainty-M}
Given the sets $E^{L_1 \land \ldots \land L_n}$ and  $E^{A \land L_1 \land \ldots \land L_n}$
that represent the extensions of the formulae  $L_1 \land \ldots \land L_n$ and   $A \land L_1 \land \ldots \land L_n$ respectively,
the \emph{degree of certainty} of $C$ in $\mathcal{M}_D$ is defined as
$\dfrac{|E^{A\land L_1 \land \ldots \land L_n}|}{|E^{L_1 \land \ldots \land L_n}|}$ 
if $|E^{L_1 \land \ldots \land L_n}| \neq 0$, $1$ otherwise.
\end{definition}

Intuitively, this definition of degree of certainty represents the conditional probability of satisfaction of the head of the integrity constraint under the assumption that the body holds in the model.

Integrity constraints annotated with a measure of their satisfaction complying with this last definition can be induced, e.g., by applying data mining techniques on a database to discover frequent relationships among its elements.
A classical approach to detect frequent patterns in a database consists in the inference of association rules~\cite{Agrawal1993,Agrawal1994}. Association rules represent assertions on predicates of the database typically denoted as implications of the form $P_1 \Rightarrow P_2$, where $P_1$ and $P_2$ are two atoms.
They are evaluated in terms of two measures of quality: the support of the association rule, which represents the extension, in the dataset, of the formula $P_1 \land P_2$, and the confidence of association the rule, which is defined as in Definition~\ref{def:degree-certainty-M} with $A=P_2$ and $L_1 = P_1$ ($n=1$). 

We emphasize that association rules do not express actual integrity constraints since they do not hold in every database model (neither do they have the same support and confidence in every dataset);  nonetheless rules with a high support and confidence provide a suitable model for query optimization in the mined database. 
Furthermore, although integrity constraints can exhibit an arbitrary structure and can involve multiple predicates, various approaches for mining more complex patterns of association rules have been also proposed in literature (e.g., multi-table extensions of association rules with conjunctive queries~\cite{Goethals2002}).

In this report we assume that integrity constraints are represented as clauses annotated with a measure of quality that is interpreted according to Definition~\ref{def:degree-certainty-M}.
We model a database as a belief logic program whose facts in the EDB hold true or false with certainty, i.e., are represented by clauses annotated by a belief interval $[1,1]$ or $[0,0]$ (but still, not every extensional predicate has necessarily a corresponding fact in the EDB, and if an atom cannot be deduced from the program, we cannot infer that its negation holds), and we extend the definition of a belief logic program to include also integrity constraints, which need not be satisfied, and are denoted by annotated rules:
\[
\begin{array}{ccc}
[v,w] & A \der L_1,\ldots, L_n  & (0\leq v \leq w \leq 1)\\
\end{array}
\]
The left value of the belief interval $[v,w]$ represents the degree of certainty of the integrity constraints. The right value is intended such that its complement $1 - w$ 
measures the frequency $\dfrac{|E_{I}^{\neg A \land L_1 \land \ldots \land L_n}|}{|E_{I}^{L_1 \land \ldots \land L_n}|}$ 
if $|E_{I}^{L_1 \land \ldots \land L_n}| \neq 0$, $0$ otherwise.

This notation follows the convention of the theory of belief logic programming of interpreting negation as explicit (or strong) negation instead of adopting the usual assumption of negation as failure. Notice that by interpreting negation as failure, the right value $w$ of the belief interval would be equal to the degree of certainty $v$ of the constraint. 

In general, data mining approaches derive association rules for which the degree of certainty is interpreted according to  the  convention of negation as failure: thus, in the following we shall consider annotated rules for which the left and right values in the interval are the same (but different annotated rules can have different belief intervals).

\subsection{Semantic Query Optimization with uncertainty}
\label{sec:SQO-combine}

The annotation of integrity constraints that expresses their degree of certainty is taken into account during the procedure of SQO to evaluate the impact of the use of this semantic knowledge on the quality of the rewriting of a query.

In this section we illustrate the technique of SQO in the presence of uncertain semantic knowledge in the simplified case of a database where the predicates occurring in the body of  the query and the integrity constraints are extensional, i.e., the IDB is empty except for the dummy rules for the queries. Notice that negated extensional predicates are allowed.
We will extend the approach to the general case of a non-empty IDB in Section~\ref{sec:approximateQAExtensions}, first with the restriction that no intensional predicate can occur negated, and then dealing with the general case where this limitation is removed.

The technique of SQO in the presence of uncertain semantic knowledge consists of the two standard phases of semantic compilation and semantic transformation.

In the first phase the residues are computed as integrity constraint fragments by applying partial subsumption between every integrity constraint and the body of dummy rules $r \der r$ or $\neg r \der \neg r$ where $r$ is an extensional predicate.

The residues are annotated with the name of the integrity constraint they are derived from.

The second phase of SQO, semantic transformation, rewrites a user's query using the residues associated with its predicates. 
Notice that we assign  to the predicates of the query residues with a multiset of annotations: the same residue can in fact be associated with the predicates of a query multiple times as the result of the application of partial subsumption between the same predicate and different integrity constraints or between different predicates and constraints.

\begin{example}\label{ex:dbBookUn}
Consider again the database introduced in Example~\ref{ex:dbBook} and assume that the integrity theory consists of the following set of uncertain integrity constraints.
\begin{align}
r_{IC_1} :  [0.8,0.8]  Authoritative(x_1, y_2) \der PCMember(x_1, x_2),  Conference(x_2, y_2, y_3)
\end{align}
``Every  program committee member of a conference on a given subject is authoritative in that subject.''
\begin{align}
r_{IC_2} : \ [0.9,0.9] \  Bestseller(y_2) \der & Authoritative(x_1, x_2),  AuthorPaper(x_1, y_2),\\
\nonumber  & Paper(y_2, x_2), x_2 = logicProgramming
\end{align}
``Every Logic Programming paper by an authoritative author in the same subject is a bestseller.''
\begin{align}
r_{IC_3} : \ [0.7,0.7] \  Bestseller(y_2) \der & Award(x_1, y_2), Paper(y_2, x_2),\\
\nonumber  & x_2 = logicProgramming
\end{align}
``Every Logic Programming paper awarded a conference prize is a bestseller.''
\begin{align}
r_{IC_4} : \ [0.8,0.8] \ x_2 \geq 20 \der Author(x_1, x_2, x_3)
\end{align}
``Every author is at least $20$ years old.''
\begin{align}
r_{IC_5} : \ [0.9,0.9] \ x_2 \geq 30 \der Author(x_1, x_2, x_3), PCMember(x_1, y_2)
\end{align}
``Every author that is a program committee member is at least $30$ years old.''

The first phase of the procedure associates with each dummy rule a corresponding set of annotated residues.\\

$
SCA_1 : Paper(o, s) \der Paper(o, s)\\
\{ (IC_2) \ Bestseller(o) \der Authoritative(x_1, s),  AuthorPaper(x_1, o), s = logicProgramming, \\
(IC_3) \ Bestseller(o) \der Award(x_1, o), s = logicProgramming \}\\
$

$
SCA_2 : Author(s, a, n) \der Author(s, a, n)\\
\{ (IC_4) \ a \geq 20 \der, \\
(IC_5) \ a \geq 30 \der  PCMember(s, x_2)\}\\
$

$
SCA_3 : AuthorPaper(s, o) \der AuthorPaper(s, o)\\
\{  (IC_2)  Bestseller(o) \der  Authoritative(s, x_2),  Paper(o, x_2),\\ x_2 = logicProgramming\}\\
$

$SCA_4 : Authoritative(a, s) \der Authoritative(a, s)\\
\{ (IC_2)  Bestseller(x_2) \der  AuthorPaper(a, x_2),  Paper(x_2, s),\\ s = logicProgramming\}\\
$

$SCA_5 : Conference(n, s, v) \der Conference(n, s, v)\\
\{ (IC_1) Authoritative(x_1, s) \der PCMember(x_1, n)
\}\\
$

$SCA_6 : PCMember(s, n) \der PCMember(s, n)\\
\{ (IC_1) Authoritative(s, x_3) \der Conference(n, x_2, x_3),\\
 (IC_5) \ x_2 \geq 30 \der Author(s, x_2, x_3)
\}\\
$

$SCA_7 : Bestseller(o) \der Bestseller(o)\\
\{\}\\
$

$SCA_8 : Award(c, o) \der Award(c, o)\\
\{ (IC_3) Bestseller(o) \der Paper(o, x_2), x_2 = logicProgramming\}\\
$

Each query to the database is adorned with the set of residues associated with its predicates, as shown, for instance, for the following three queries $Q_1$ in Equation~\eqref{eq:Q1}, $Q_2$ in Equation~\eqref{eq:Q2} and $Q_3$ in Equation~\eqref{eq:Q3} to the database (all variables are output variables).\\

\noindent $Q_1$: ``Find program committee members of conferences in Australia."
\begin{align}\label{eq:Q1}
Q_1: q_1(x_1, x_2, y_2, y_3) \der & PCMember(x_1, x_2), Conference(x_2, y_2, y_3),\\
 \notag & y_3 = Australia
\end{align}
\[
\begin{array}{lll}
Q_1: \der & PCMember(x_1, x_2) &\{ 
 (IC_1) \  Authoritative(x_1, z_2) \der Conference(x_2, z_2, z_3),\\
&&  (IC_5) \ z_2 \geq 30 \der Author(x_1, z_2, z_3)\}\\
 & Conference(x_2, y_2, y_3) &\{ (IC_1) \ Authoritative(t_1, y_2) \der PCMember(t_1, x_2) \},\\
 & y_3 = Australia &
\end{array}
\]
After applying again subsumption between the residues of a predicate and the rest of the query, and removing redundant residues obtained from the same integrity constraint, the set can be rewritten as:
\[
\begin{array}{l}
 \{ 
 (IC_1) \  Authoritative(x_1, y_2) \der,\\
 (IC_5) \ z_2 \geq 30 \der Author(x_1, z_2, z_3)\}
\end{array}
\]

\noindent The second query is:
 
\noindent $Q_2$: ``Find authoritative authors in Logic Programming who wrote an awarded Logic Programming paper that is a bestseller."
\begin{align}\label{eq:Q2}
Q_2: q_2(x_1, x_2, y_2, z_2) \der & Authoritative(x_1, x_2), AuthorPaper(x_1, y_2), Paper(y_2, x_2)\\
\notag	 & Award(z_2, y_2), Bestseller(y_2), x_2 = logicProgramming
	\end{align}
\[
\begin{array}{llll}
Q_2: \der 
& Authoritative(x_1, x_2) &\{(IC_2) \ Bestseller(t_2) \der & AuthorPaper(x_1, t_2),\\
		                   & && Paper(t_2, x_2),\\
		                   &&& x_2 = logicProgramming\},\\
& AuthorPaper(x_1, y_2)   &\{(IC_2) \ Bestseller(y_2) \der & Authoritative(x_1, u_2), \\
						   & & & Paper(y_2, u_2),\\
						   &&& u_2 = logicProgramming \},\\
& Paper(y_2, x_2) 	     &\{(IC_2) \ Bestseller(y_2) \der& Authoritative(v_1, x_2),\\
						&&& AuthorPaper(v_1, y_2),\\ 
						&&& x_2 = logicProgramming,\\
						   & &(IC_3) \ Bestseller(y_2) \der & Award(q_1, y_2),\\
						   &&& x_2 = logicProgramming  \},\\
& Award(z_2, y_2) 	     &\{ (IC_3)\ Bestseller(y_2) \der & Paper(y_2, r_2),\\
&&& r_2 = logicProgramming \},\\
& Bestseller(y_2)	     &\{ \},&\\	   
& x_2 = logicProgramming &&
\end{array}
\]
Again, the set can be rewritten as:\\
$
 \{ 
 (IC_2;IC_3) \  Bestseller(y_2) \der \}\\
$

\noindent The third query is:

\noindent $Q_3$: ``Find program committee members that are Spanish authors."
\begin{align}\label{eq:Q3}
Q_3: q_1(x_1, x_2, y_2, y_3) \der & PCMember(x_1, x_2), Author(x_1, y_2, y_3), y_3 = Spanish
\end{align}
\[
\begin{array}{lll}
Q_3: \der & PCMember(x_1, x_2) &\{ 
(IC_1) Authoritative(x_1, z_2) \der Conference(x_2, z_2, z_3),\\
&& (IC_5) \ t_2 \geq 30 \der Author(x_1, t_2, t_3) \},\\
 & Author(x_1, y_2, y_3), &\{ (IC_4) \ y_2 \geq 20 \der, \\
 & &(IC_5) \ y_2 \geq 30 \der PCMember(x_1, u_2) \},\\
 & y_3 = Spanish &
\end{array}
\]
\noindent and the set can be rewritten as:
\[
\begin{array}{l}
 \{ 
 (IC_1) Authoritative(x_1, z_2) \der Conference(x_2, z_2, z_3),\\
 (IC_5) \ y_2 \geq 30 \der,\\
 (IC_4) \ y_2 \geq 20 \der\}
\end{array}
\]
\hfil\qed

\end{example}

If the residues were derived from certain integrity constraints, as in the classical procedure, the most effective execution plan could then be selected for the transformation of a query without altering the set of answers retrieved from the database.
However, in the presence of uncertain semantic knowledge, the use of residues to rewrite the query may not guarantee the semantic equivalence with the original one. Hence, the choice of a rewriting must take into account 
two factors: its cost of execution, which depends on the cardinality of the set of terms satisfying the predicates in the database and the availability of indexes on their attributes, and  the accuracy  of the answers to the transformed query.

The accuracy of approaches that yield approximate answers is usually evaluated -- a posteriori -- in terms of precision and recall. 
Precision measures the fraction of tuples in the retrieved data that are correct answers to the original query.
Recall measures the fraction of answers to the original query that are actually returned after processing the rewritten query.

Herein, we  assess the quality of a candidate rewriting by estimating measures of correctness and completeness (presented in the next paragraphs) of the answers to the new query, evaluating the annotation of certainty associated with the residues used for the transformation.

\begin{remark}
Notice that, as mentioned in the introduction, the residues can be obtained from non-independent sources of uncertain semantic knowledge: for instance, in the database of Example~\ref{ex:dbBookUn} the bodies of both the integrity  constraints $IC_2$ and $IC_3$ that derive the residue $Bestseller(y_2)$ rely on the same atom $Paper(y_2, logicProgramming)$.

The presence of uncertain and correlated information raises the  issue  of how to properly combine this knowledge when  we look for an evaluation of the accuracy of the overall conclusions that can be derived from these sources.
Assume that we want to determine the degree of certainty of the atom $Bestseller(y_2)$ to assess whether it is suitable to use it to rewrite the query $Q_2$ into the simplified form 
$\der  Authoritative(x_1, x_2),$ $ AuthorPaper(x_1, y_2),$ $ Paper(y_2, x_2), Award(z_2, y_2), $ $x_2 = logicProgramming$ where the elimination of the predicate allows us to remove a join.

If the constraints were simply annotated with a degree of certainty $v_i$ ($i=1,2$) expressing the likelihood by which the head holds true given the truth of the atoms in the body, the certainty of the atom $Bestseller(y_2)$ could not be computed by straightforwardly combining the two degrees as if the constraints were independent (the value asserted for $Bestseller(y_2)$ would be too high).
However, this is the path typically followed by the classical approaches in the field of quantitative reasoning that deal with uncertainty and inconsistency in knowledge representation, which mostly disregard correlation of evidence obtained from non-independent sources. Belief Logic Programming differs from them in that it provides an appropriate theory for the combination of evidence derived from such uncertain elements of information .  

We relied on the theory of BLP and modeled the database as a belief logic program and the integrity constraints as annotated rules associated with a belief interval, purposely to apply the results of this theory for the analysis and combination of evidence derived from the available, uncertain and possibly correlated, knowledge.

\end{remark}

In the following, we first illustrate the evaluation of the accuracy of the rewritings, by describing the impact on correctness and completeness of the possible types of transformation of a query.
Next, we present the pseudo-code of the second phase of the procedure (semantic transformation), which chooses a rewriting for a given query on the basis of its cost and these measures of accuracy.

Once a rewriting has been selected and the rewritten query is executed on the database, we can evaluate a more refined estimate of correctness of a concrete answer returned to the user.

\subsubsection{Correctness}
\label{sec:correctness}

\begin{definition}\label{def:correctness}
Let $D =\langle EDB, IDB, IC \rangle$ be a database and let $b$ be the belief logic program associated with $D$ and extended with the integrity constraints in $IC$.

For a query $Q: q(\vec{z}) \der F(\vec{y})$ and its rewriting $Q'$, let $\vec{x}$ be an answer to $Q'$. The correctness of the answer is measured by the belief by $b$ in $q(\vec{x})$.
\end{definition}

\begin{remark}
Remind that for each query $Q: q(\vec{z}) \der F(\vec{y})$ we define a fictitious rule 
$
r_G: \ [1,1] $ $q(\vec{z}) \der F(\vec{y})
$.

\end{remark}

The correctness of the transformation of semantic query optimization is considered at the level of each single tuple in the set of answers and is measured by the ``likelihood'' that a given answer to  the rewritten query $Q'$ is an answer to the original query.

In general, this value of belief is not necessarily the same for different answers to the rewritten query. We are thus interested in finding a lower bound on the possible value of belief for a generic answer, and then choose a suitable rewriting that does not yield an excessive impairment of the answers returned by the procedure.

To illustrate the evaluation of the correctness of an answer retrieved after the execution of a transformed query, we examine three basic examples of rewriting, where a query is optimized either by  introducing in its body  a new atom (an extensional atom or a restriction), or by removing an extensional atom; thereafter we deal with the general transformation of a query to which one or more of these optimizations has been applied.

\begin{description}
\item [Introduction of an atom]
We first consider a rewriting that introduces an extensional atom $A$ in the body of the query.

Let $IC_1,\ldots,IC_k$ be the set of integrity constraints, with belief interval $[v_i,v_{i}]$ ($1 \leq i \leq k$) respectively, each having as head the same atom $A$.

Consider a query $Q: q(\vec{z})  \der F(\vec{y})$ such that $F(\vec{y})$ is subsumed by the body of some among $IC_1,\ldots,IC_k$ and thus has an associated residue $A \der$ obtained from these constraints, and let $\vec{a}$ be an answer to the query $Q': q'(\vec{z})\der F(\vec{y}), A$ obtained from $Q$ by adding the atom denoted by the residue. 

Clearly, each substitution $\sigma$,  with $dom(\sigma) = \vec{y}$ and  $\vec{z} \sigma = \vec{a}$, which satisfies $\models (F(\vec{y}) \land A) \sigma$  is also a substitution satisfying $\models F(\vec{y})  \sigma$, and $\vec{a}$ is an answer to $Q$ with belief $belief(q(\vec{a}))$ equal to $1$.

The same result holds also in the case that a restriction is introduced in the body of the query or more than one atoms/restrictions are introduced.

\begin{example}
For instance, query $Q_1$ of Example~\ref{ex:dbBookUn} can be rewritten as $Q'_1: q'_1(x_1, x_2, y_2, y_3) \der PCMember(x_1, x_2), Conference(x_2, y_2, y_3), Authoritative(x_1, $ $y_2), y_3 = Australia$ by introducing the atom $Authoritative(x_1, y_2)$.
Clearly, every answer to the rewritten query is an answer to the original one.
\end{example}

\item [Removal of an atom] We now deal with the case of a rewriting that removes one or more atoms from the body of a query.

Let $p_i/n_i$ ($1\leq i\leq m$) be predicates occurring as head of the integrity constraints in the sets $P_i = \{ IC_1^i, \ldots IC_{k_i}^i  \}$ respectively.
Consider a query $Q: q(\vec{z}) \der F(\vec{y})$ whose body is subsumed by the body of some integrity constraints in each set $P_i$, say $IC_1^i, \ldots IC_{h_i}^i$ ($1 \leq h_i \leq k_i$) for all $i$. Thus, $Q$ has $\{ p_1(\vec{x_1}) \der, \ldots,  p_m(\vec{x_m}) \der \}$
 as residues.
 
Assume that atoms  $p_i(\vec{x_i})$ occur in $Q$, i.e., $F(\vec{y})$ is $G(\vec{y}) \land \bigwedge_{i=1}^{m} p_i(\vec{x_i})$ for some atomic formula $G(\vec{y})$, and consider the rewriting of $Q$ as the query $Q': q'(\vec{z})\der G(\vec{y})$ where all predicates $p_i$ have been eliminated.

We consider first the simplified case that all the variables of the query are output variables, i.e., $\vec{y} = \vec{z}$. At the end of this paragraph we will remove this limitation to deal with the general case that $\vec{z}$ is a possibly proper subarray of $\vec{y}$.

Let $\vec{a}$ be an answer to $Q'$, i.e., $\models G(\vec{a})$, and denote by  $\sigma$ the substitution $\{\vec{y}/\vec{a}\}$.

Following the notation of the theory of BLP, to compute the value of belief in $q(\vec{a})$ we can build a suitable proof DAG. The DAG has an atom $q(\vec{a})$ as root, reached by an edge labeled $[1,1]$ from an r-node $r_G(\sigma)$ having as ground children the literals in $G(\vec{y}) \sigma$ and the atoms $p_i(\vec{x_i}) \sigma$ ($1 \leq i \leq m$). Each $p_i(\vec{x_i})\sigma$ is the root of a subgraph of height $2$ having as first level the rule nodes corresponding to the constraints $IC_1^i,\ldots,IC^i_{k_i}$ and, as their children, at the second level, the atoms in the body of such constraints.
For a concrete answer $\vec{a}$, the belief in $q(\vec{a})$ by the belief logic program extended with the integrity constraints --in the interpretation where $G(\vec{a})$ holds true with certainty-- could then be computed by applying the BLP query answering algorithm, where also the rule edges in the proof DAG corresponding to the constraints are taken into account.

Notice that the DAG does not include the r-node (and the corresponding departing edge) for the extensional atoms that have been removed. In fact, in the presence of these facts, the lowest bound on the value of correctness corresponds to the case where not all the atoms that have been eliminated are true, and thus is necessarily equal to 0. To compute a meaningful value, instead, we determine the lower bound  considering the belief in $q(\vec{a})$ by the belief logic program that is extended with the integrity constraints and does not include the facts denoting the removed predicates.

\begin{example}\label{ex:removalCorrectness}
Consider again Example~\ref{ex:dbBookUn}.
Assume that the EDB of the database includes, among others, the following facts.

\noindent E1: Paper($oid_1$, logicProgramming) , which has as title ``The Stable Model Semantics for Logic Programming''\\
E2: Author(Gelfond, 69, Russian)\\
E3: Author(Lifschitz, 67, Russian)\\
E4: AuthorPaper(Gelfond, $oid_1$)\\
E5: AuthorPaper(Lifschitz, $oid_1$)\\
E6: Authoritative(Gelfond, logicProgramming)\\
E7: Authoritative(Lifschitz, logicProgramming)\\
E8: PCMember(Gelfond, ICLP2004)\\
E9: Bestseller($oid_1$)\\
E10: Award(ICLP2004, $oid_1$)\\
E10: Conference(ICLP2004, logicProgramming, StMalo)\\

Assume also that there is another integrity constraint:
\begin{align*}
r_{IC_6} : \ [0.6,0.6] \  Bestseller(y_2) \der & PCMember(x_1, x_2), AuthorPaper(x_1, y_2)
\end{align*}
``A program committee member writes only works that are bestsellers.''\\

Consider the query $Q_2$, reported again below. 
\begin{align*}
Q_2: q(x_1, x_2, y_2, z_2) \der  & Authoritative(x_1, x_2), AuthorPaper(x_1, y_2), Paper(y_2, x_2),\\
 \notag &  Award(z_2, y_2), Bestseller(y_2), x_2 = logicProgramming
\end{align*}

The set of integrity constraint that have predicate $Bestseller$ as head is $P = \{IC_2, IC_3, IC_6\}$, but only $IC_2$ and $IC_3$ generate a useful residue for the query. The query can be rewritten as 
\begin{align*}
Q_2': q'(x_1, x_2, y_2, z_2) \der & Authoritative(x_1, x_2), AuthorPaper(x_1, y_2), Paper(y_2, x_2), \\
\nonumber	 & Award(z_2, y_2), x_2 = logicProgramming
\end{align*}
by removing the atom $Bestseller(y_2)$ using the available residue.

$Q_2'$ has the answer $\vec{a} = (Gelfond, logicProgramming, oid_1, ICLP2004)$.
To compute the belief by the blp in $q(Gelfond, logicProgramming, oid_1, ICLP2004)$, we consider the aforementioned proof DAG, shown in Figure~\ref{fig:DAGRemovalCorr}.

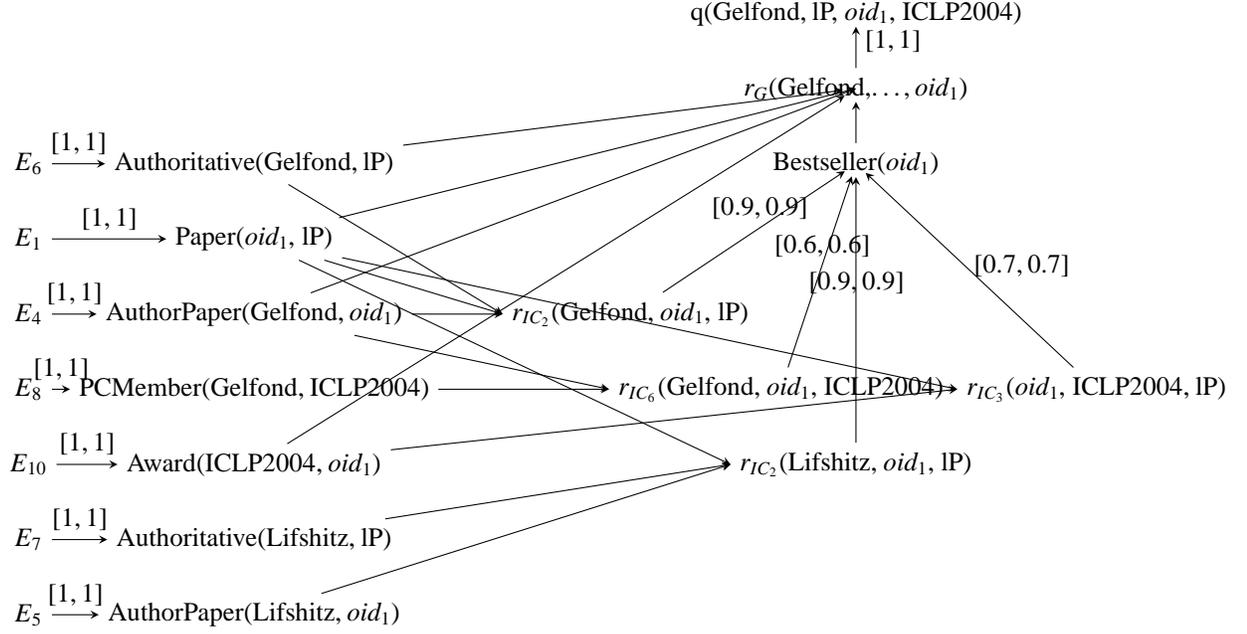
\begin{figure}[h!]
\begin{tikzpicture}[->, >=stealth,auto,%
    node distance=1cm,%
    every stat/.style={draw=none,text=black, minimum size, inner sep=0pt},%
    every edge/.style={draw,solid}%
    ]
    
  \node         (Au)                    {Authoritative(Gelfond, lP)};
  \node         (P) [below of= Au]       {Paper($oid_1$, lP)};  
  \node         (Ap) [below of= P] {AuthorPaper(Gelfond, $oid_1$)};
  \node         (Pc) [below of= Ap]  {PCMember(Gelfond, ICLP2004)};  
  \node         (Aw) [below of= Pc]  {Award(ICLP2004, $oid_1$)};
  \node         (Au2) [below of= Aw]     {Authoritative(Lifshitz, lP)};
  \node         (Ap2) [below of= Au2]     {AuthorPaper(Lifshitz, $oid_1$)};
  \node         (B) [right of= Au, xshift=7cm]     {Bestseller($oid_1$)};  
  \node         (rg) [above of= B]     {$r_G$(Gelfond,$\ldots,oid_1)$};
  \node         (q) [above of= rg]     {q(Gelfond, lP, $oid_1$, ICLP2004)};  
  \node         (ic2b) [below of= B, yshift=-3cm]     {$r_{IC_2}$(Lifshitz, $oid_1$, lP)};

  \node         (ic3) [right  of= ic2b, yshift=+1cm, xshift=2.2cm]     {$r_{IC_3}$($oid_1$, ICLP2004, lP)};
  \node         (ic6) [left of= ic2b, yshift=+1cm]     {$r_{IC_6}$(Gelfond, $oid_1$, ICLP2004)};
  \node         (ic2a) [above of= ic6,  xshift=-2cm]     {$r_{IC_2}$(Gelfond, $oid_1$, lP)};    

  \node         (eAu) [left of= Au, xshift=-2cm]     {$E_6$}; 
  \node         (eAp) [left of= Ap, xshift=-2cm]     {$E_4$}; 
  \node         (eP) [left of= P, xshift=-2cm]     {$E_1$};
  \node         (eAw) [left of= Aw, xshift=-2cm]     {$E_{10}$};    

  \node         (ePc) [left of= Pc, xshift=-2cm]     {$E_8$}; 
  \node         (eAu2) [left of= Au2, xshift=-2cm]     {$E_7$}; 
  \node         (eAp2) [left of= Ap2, xshift=-2cm]     {$E_5$};

  \path[->] (Au)  edge[shorten >= 5pt]   (rg.center)
				  edge	(ic2a.west)
			(Ap)   edge  (rg.center)
				  edge  (ic2a.west)
				  edge  (ic6.west)
			(P)   edge  (rg.center)
				  edge  (ic2a.west)
				  edge  (ic2b.west)				  
				  edge  (ic3.west)
			(Aw)   edge[shorten >= 5pt]  (rg.center)
				  edge  (ic3.west)
			(Pc)   edge  (ic6.west)
			(Au2)  edge  (ic2b.west)
			(Ap2) edge  (ic2b.west)
			(B)   edge[shorten >= 5pt]  (rg.center)
			(rg)   edge[shorten >= 5pt]  node[right] {$[1,1]$} (q.center)				  
			(ic2a)   edge[shorten >= 5pt] node[above] {$[0.9,0.9]$}  (B.center)
			(ic2b)   edge[shorten >= 5pt] node[above] {$[0.9,0.9]$} (B.center)
			(ic3)   edge[shorten >= 5pt] node[right] {$[0.7,0.7]$} (B.center)
			(ic6)   edge[shorten >= 5pt] node[above] {$[0.6,0.6]$} (B.center)  		
			(eAu) edge node[above] {$[1,1]$}  (Au)
			(eAp) edge  node[above] {$[1,1]$} (Ap)
			(eP) edge  node[above] {$[1,1]$}  (P)
			(eAw) edge node[above] {$[1,1]$}  (Aw)	
			(eAu2) edge node[above] {$[1,1]$}  (Au2)
			(eAp2) edge  node[above] {$[1,1]$} (Ap2)
			(ePc) edge  node[above] {$[1,1]$}  (Pc)													     
;   
\end{tikzpicture}
\caption{Proof DAG to compute a lower bound on the belief of $q(\vec{a})$ in Example~\ref{ex:removalCorrectness} Constant $logicProgramming$ is abbreviated as $lP$.\label{fig:DAGRemovalCorr}}
\end{figure}

Notice that the atoms $Authoritative(Lifschitz, logicProgramming),$   $AuthorPaper$ $(Lifschitz,oid_1),$ $ PCMember(Gelfond, ICLP2004) $ contribute to the computation of the combined belief of $Bestseller(oid_1)$, although they do not occur in the body of the query grounded by the substitution $\sigma$.
\hfil\qed
\end{example}

The exact value of the belief can be computed only by knowing the combined belief and disbelief (i.e., the likelihood) associated with all the atoms in the DAG, both the literals in $G(\vec{a})$, which are known to be true with certainty, and the literals in the body of all the possible ground instantiations of the integrity constraints having predicates $p_i(\vec{x_i})$ as head, whose truth value may differ for different answers to the rewritten query.

When evaluating the belief for a generic answer, however, the contribution provided by these annotated rules to the value of belief cannot be naively discarded, even to determine an under approximation of this value, since BLP semantics is non-monotonic (thus the introduction of new sources of knowledge to a set of annotated rules deriving a given atom can impact on the computation of its combined belief, possibly increasing or reducing it).

Also, a set of instantiations of the body of one such integrity constraint can contribute to the support of a truth valuation with an interval that, depending on the chosen combination function and the size of the set, can remain bounded between the extremes of the belief interval of the constraint or can also reach limit values of $1$ or the trivial lower bound of $0$.
For instance, applying Dempster's combination function to a multiset of belief intervals $S= \{[v,v], \ldots, [v,v]\}$, the resulting combined belief factor $\phi^{DS}(\{[v,v], \ldots, [v,v]\})$ tends to approach $[0,0]$ (resp. $[1,1]$) for $0 < v < 0.5$ (resp. for $0.5 < v < 1$) as the cardinality of the multiset of belief intervals increases.

A viable approach consists in computing a lower bound on the value of belief by introducing some assumptions on the combination functions $\phi$ used to determine the beliefs.

We shall consider combination functions $\phi$ such that for every $0 \leq v_1 \leq 1$, $0 \leq v_2\leq 1$ and a multiset $S = \{[v_2, v_2], [v_2, v_2], \ldots, [v_2, v_2]\}$ of belief intervals $[v_2, v_2]$ of any cardinality, $\phi_{V}([v_1,v_1], S) \geq \phi_{V}([v_1,v_1], [v_2, v_2])$.
Under this hypothesis on $\phi$, every information derived from multiple different instantiations of the same constraint will increase or leave unchanged the overall belief on the atom it has as head.
Thus, as regards the integrity constraints that derive residues that have a removed atom as head but that do not subsume the body of the query, we cannot know a priori whether their body is true and their contribution  to the combined belief of the removed atom must thus be taken into account, but we can at most  consider the contribution provided by only one of the possible instantiations for the integrity constraint.

Among the classical combination functions, this assumption is satisfied by $\phi^{MAX}$, $\phi^{MIN}$.
In fact, clearly, 
$\phi^{MAX}([v_1,v_1], S) = \phi^{MAX}([v_1,v_1], \phi^{MAX}(S)) =$ $ \phi^{MAX}([v_1, v_1],$ $ [v_2, v_2])$, and equivalently,
$\phi^{MIN}([v_1,v_1], S) = \phi^{MIN}([v_1,v_1], \phi^{MIN}(S)) =$ $ \phi^{MIN}$ $([v_1, v_1], [v_2, v_2])$.

However, it is not satisfied in general by Dempster's combination rule $\phi^{DS}$.
A sufficient condition that guarantees that it is valid also for $\phi^{DS}$ is that $v_1, v_2 \geq 0.5$.
\begin{enumerate}
\item \label{item:phiDS1} We first show that, for every $0 \leq v_1 \leq 1$, $0 \leq v_2\leq 1$, $\phi^{DS}([v_1,v_1], [v_2, v_2]) = [v,w]$ with $w= v$ and $v \geq max(v_1, v_2)$ iff $v_1, v_2 \geq 0.5$.
By definition, it holds for the trivial assignment $[v_1,v_1]=[0,0]$ and $[v_2,v_2]=[1,1]$.
For the other possible values of the interval, $ v = w = \frac{v_1 v_2}{1 + 2 v_1 v_2 - v_1 - v_2}$. Since $v \geq v_2$ iff $v_1 \geq 0.5$, (and conversely, $v \geq v_1$ iff $v_2 \geq 0.5$), the condition holds.

\item \label{item:phiDS2} Also, for every $0 \leq v \leq 1$, $0 \leq v_1\leq 1, 0 \leq v_2 \leq 1$, $\phi^{DS}_{V}([v_1,v_1], [v_2,v_2]) \leq \phi^{DS}_{V}([v_1,v_1], [v,v])$ iff $v_2 \leq v$.
By definition,  in fact, 
$\frac{v_1 v_2}{1 + 2 v_1 v_2 - v_1 - v_2} \leq \frac{v_1 v}{1 + 2 v_1 v - v_1 - v}$ iff
 $v_2 \leq v$.

\item Then, by point~\ref{item:phiDS1}, if $v_1, v_2 \geq 0.5$, we have that $\phi^{DS}([v_1,v_1], S) = \phi^{DS}([v_1,v_1],$ $ \phi^{DS}(S)) = \phi^{DS}([v_1,v_1], [v,v])$ with $v \geq v_2$.
By point~\ref{item:phiDS2}, $\phi^{DS}([v_1,v_1], [v,v]) \geq \phi^{DS}([v_1,v_1], [v_2,v_2])$.
 
\end{enumerate}
In the sequel we shall assume that all the integrity constraints have been derived or mined for frequent patterns in the database, such that their belief interval is $[v,v]$ with $v\geq 0.5$, and thus the condition is satisfied.

Then, we compute a lower bound on the likelihood by which an answer $\vec{a}$ returned after processing $Q'$ can be considered as an answer to the original query.

Consider the previous DAG, where we do not explore the graph below its third level of a-nodes from the root. In principle, we should enumerate the interpretations where the literals in $G(\vec{a})$ and $p_i(\vec{x_i}) \sigma \ \forall 1 \leq i \leq m$ are true, and the other a-nodes in the bodies of the rules or constraints deriving $p_i(\vec{x_i})$ can have any truth value. Let $q_1, \ldots, q_f$ denote the a-nodes of the graph different from those corresponding to the eliminated predicates $p_i$: according to the theory of BLP, the support of a truth valuation $I$ is 
$\prod_{j=1}^{f} \alpha_j \prod_{i=1}^{m} Val(\phi(R_{p_i}), I(p_i(\vec{x_i})\sigma)) = \prod_{j=1}^{f} \alpha_j \prod_{i=1}^{m} \phi_{V}(R_{p_i})$,
where, for $q_j$ with combined belief $v_j$ and combined disbelief $w_j$,  $\alpha_j$ is $v_j$ if $q_j$ is true (or $1-w_j$ if $q_j$ is false, or $w_j - v_j$ if $q_j$ is uncertain, respectively),
and $R_{p_i}$ is the multiset of belief intervals of the annotated rules deriving $p_i(\vec{x_i}) \sigma$ whose body is true in the interpretation.
The belief in $q(\vec{a})$ corresponds to the sum of supports of the truth valuations where $q(\vec{a})$ is true.
A lower bound on this value could be obtained by selecting assignments of values $0$ or $1$ to the combined belief $v_j$ and disbelief $w_j$ of the a-nodes $q_j$ ($1 \leq j \leq f$) yielding an interpretation with minimum support, i.e. that minimizes the possible products $\prod_{i=1}^{m} \phi_{V}(R_{p_i})$.

Given the above assumption on the belief combination function, we can compute a lower bound by considering the truth valuation where the assignment of truth values sets as true with certainty
 the ground bodies of the integrity constraints that subsume the query $\der G(\vec{a})$ and, for each constraint that derives residues that have a removed atom as head but that do not subsume the body of the query, sets as true at most one of its possible instantiations The possible instantiations of constraints, among these, that are set as true are chosen as those that minimize the above product; otherwise they are discarded. 
 
 Notice that, however, the choice of which integrity constraint to consider, for the computation of the combined belief of the removed atoms, can be greatly simplified if we consider the three usual belief functions $\phi^{MAX}, \phi^{MIN}, \phi^{DS}$.
 In fact, as regards Dempster's combination rule, by point~\ref{item:phiDS1}, if a predicate $p_i(\vec{x_i}) \sigma$ has been removed since derived by a set of integrity constraints whose body subsumes the query $\der G(\vec{a})$, any further contribution derived from other integrity constraints will increase or leave unchanged the overall belief in the atom (and in particular the contribution of the integrity constraints that do not subsume the body of the query can be discarded without affecting the lower bound).
 
The same holds for $\phi^{MAX}$, by definition.
Thus, we can compute a lower bound as $belief(q(\vec{a})) \geq \prod_{i=1}^{m} \phi_{V}(Q_{p_i})$ , where $Q_{p_i}$ is the multiset of belief intervals of annotated rules having $p_i(\vec{x_i}) \sigma$ as head whose  body is composed only of predicates in $ G(\vec{a})$.

Concerning $\phi^{MIN}$, it suffices to choose, among the residues that derive an atom that has been removed, the one having the lowest belief interval and to set as true its body. In this way, however, we do not take into account the fact that the possible instantiations of these constraints can derive different removed atoms, with different belief intervals, and thus we should choose an assignment of truth values that minimizes all the possible products and is consistent: in any case, in this way we find a correct lower bound (though not strict).

For simplicity, in the sequel of this report we will refer to any of the first two cases (i.e., $\phi^{MAX}$ and $\phi^{DS}$). The case for $\phi^{MIN}$ or a generic combination function satisfying the previous assumption can be handled as previously mentioned, by selecting the integrity constraints that minimize the combined belief in the removed atoms.

\begin{example}
For the previous example, discarding the contribution of atom  $PCMember(Gelfond, ICLP2004)$ (and also of atoms $AuthorPaper$ $(Lifschitz,$ $oid_1)$,  $Authoritative(Lifschitz, logicProgramming)$), we can compute a lower bound on the belief as $l = \phi^{DS}([v_2, w_2], [v_3, w_3]) = \phi^{DS}([0.9, 0.9], [0.7, 0.7])$.
\end{example}

We now turn to the general case where not all the variables of the query are output variables.
 
Let $\vec{a}$ be an answer to $Q'$. 
There can be multiple substitutions $\sigma_j$ ($1 \leq j \leq n$), with $dom(\sigma_j) = \vec{y}$ and $\vec{z} \sigma_j = \vec{a}$, such that $\models  G(\vec{y}) \sigma_j$. 
This case, however, can be reduced to the basic one where there is only one such substitution $\sigma$. In fact, for any belief function $\phi$ such that $\phi([1,1],[1,1] = [1,1])$, the belief in $q(\vec{a})$ considering only the contribution of a valid substitution is not greater than the value computed considering all the valid substitutions $\sigma_1, \ldots, \sigma_n$.

\item [General rewriting]
We now examine a general rewriting of a query $Q: q(\vec{z})  \der F(\vec{y})$ rewritten by applying one or more of the previous basic transformations. 

If the rewritten query $Q': q'(\vec{z})\der G(\vec{y})$ has been obtained from $Q$ without removing any predicate, then each of its answers $\vec{a}$ is valid also for the query $Q$ and $belief(q(\vec{a})) = 1$.

Otherwise, at least an atom has been removed from $Q'$ w.r.t. the original query.

Let $\vec{a}$ be an answer to $Q'$ and, as in the previous case of removal of a predicate, let $\sigma$ be any substitution for the variables of $Q$ with $dom(\sigma) = \vec{y}$ and $\vec{z} \sigma = \vec{a}$, such that $\models  G(\vec{y}) \sigma$. 

To compute a lower bound on the belief in $q(\vec{a})$ we build a proof DAG, in a similar way as in the case of the removal of an extensional atom:
\begin{enumerate}
\item The graph has root $q(\vec{a})$, reached by an edge labeled $[1,1]$ from an r-node $r_G(\sigma)$
\item The first level of a-nodes includes the atoms in the body of $Q$
\item \label{item:correctnessGeneral} For each extensional atom in $Q$ that has been removed from $Q'$, 
\begin{enumerate}
\item \label{item:correctnessGeneral1} include in the proof DAG the subtrees corresponding to the integrity constraints used to derive it.
For every atom in the body of these constraints:
\begin{enumerate}
\item either it occurs in the body of $Q'$ (and thus holds true with certainty) 
\item or it is an extensional atom removed from the body of $Q$. If we assume that an extensional atom is no more removed once it has been introduced in the rewritten query, then the contribution of this atom is already taken into account in step~\ref{item:correctnessGeneral}.
\end{enumerate}
\end{enumerate}
\end{enumerate}

Under the same assumption on $\phi$ and the belief intervals of the integrity constraints as before, a lower bound can be computed exactly as in the base case by setting to false all the formulae corresponding to bodies of  constraints for which it is not explicitly known a combined belief.

\begin{example}\label{ex:generalCorrectness}
Consider a query
 
\noindent $Q_4$: ``Find program committee members of a conference on Logic Programming who wrote a paper in Logic Programming that is a bestseller; return author's surname and oid of the paper."
\begin{align}\label{eq:Q4}
Q_4: q_4(x_1,y_2)\der & PCMember(x_1, x_2), AuthorPaper(x_1, y_2),\\
\nonumber    & Paper(y_2, u_2), Bestseller(y_2), \\
\nonumber	 & Conference(x_2, u_2, v_3), u_2 = logicProgramming
\end{align}
	
Assuming that the predicate $Authoritative$ has a small extension, while the extension of $Bestseller$ is large, it is convenient to rewrite the query by exploiting the corresponding residues (not shown here for brevity) introducing an atom $Authoritative$ and removing $Bestseller$.
\begin{align*}
Q'_4: q'_4(x_1, y_2) \der & Authoritative(x_1, u_2), PCMember(x_1, x_2),\\
\nonumber			 & AuthorPaper(x_1, y_2), Paper(y_2, u_2),\\
\nonumber	 & Conference(x_2, u_2, v_3), u_2 = logicProgramming
\end{align*}

For instance, the rewritten query has the answer $\vec{a} = (Gelfond, oid_1) $.

Figure~\ref{fig:DAGGeneralCorr} shows the proof DAG on which we could compute a lower bound on the belief	in $q(Gelfond, oid_1)$, given by  0.9.

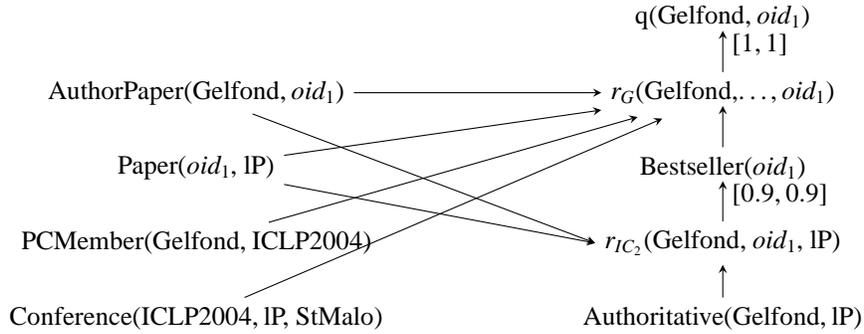
\begin{figure}[h!]
\centering
\begin{tikzpicture}[->, >=stealth,auto,%
    node distance=1cm,%
    every stat/.style={draw=none,text=black, minimum size, inner sep=0pt},%
    every edge/.style={draw,solid}%
    ]    

  \node         (Ap) [] {AuthorPaper(Gelfond, $oid_1$)};
  \node         (P) [below of= Ap]       {Paper($oid_1$, lP)}; 
  \node         (Pc) [below of= P]  {PCMember(Gelfond, ICLP2004)};
  \node         (C) [below of= Pc]     {Conference(ICLP2004, lP, StMalo)};       
  \node         (B) [right of= P, xshift=6cm]     {Bestseller($oid_1$)};  
  \node         (ic2) [below of= B]     {$r_{IC_2}$(Gelfond, $oid_1$, lP)};
  \node         (At) [below of= ic2] {Authoritative(Gelfond, lP)};
  \node         (rg) [above of= B]     {$r_G$(Gelfond,$\ldots,oid_1)$};
  \node         (q) [above of= rg]     {q(Gelfond, $oid_1$)};  

  \path[->] (Pc)  edge[shorten >= 5pt]   (rg)
			(C)   edge[shorten >= 5pt]   (rg)
			(Ap)  edge  (rg)
				  edge  (ic2.west)
			(P)   edge  (rg)
				  edge  (ic2.west)
			(B)  edge[shorten >= 5pt]  (rg.center)
			(At)  edge  (ic2)
			(rg)   edge[shorten >= 5pt]  node[right] {$[1,1]$} (q.center)				  
			(ic2)   edge[shorten >= 5pt] node[right] {$[0.9,0.9]$} (B.center)
;   
\end{tikzpicture}
\caption{Proof DAG to compute a lower bound on the belief of $q(\vec{a})$ in Example~\ref{ex:generalCorrectness}.\label{fig:DAGGeneralCorr}}
\end{figure}
\hfill\qed
\end{example}
\end{description}

Notice that, as previously mentioned, this measure of correctness can also be evaluated after executing the query to estimate the accuracy of an answer actually retrieved by the procedure and, thus, to provide the users with an estimate of the quality of the answers returned to them.
Clearly, once a query has been processed, we can compute a stricter lower bound on correctness, since we know which is the truth value of the possible ground instantiations of the body integrity constraints used to rewrite the query. Thus, for instance, for the query in Example~\ref{ex:removalCorrectness}, the belief can be computed taking into account also the contribution of atoms $Authoritative(Lifschitz, logicProgramming),$   $AuthorPaper(Lifschitz,oid_1)$.
In general, for an answer $\vec{a}$, all the substitutions $\sigma_i$ for the variables of $Q$ with $dom(\sigma_i) = \vec{y}$ and $\vec{z} \sigma_i = \vec{a}$ such that $\models  G(\vec{y}) \sigma_i$ would be known and can be thus evaluated to determine the belief. Only the body of the integrity constraints that do not produce useful residues for the query is not necessarily known to be true (as $PCMember(Gelfond, ICLP2004)$ in Example~\ref{ex:removalCorrectness}), but computing the combined belief and disbelief of the atoms in their  body would require executing corresponding queries on the database, with clear loss of performance.

The correctness of a concrete answer can be estimated building the previously described  proof DAGs, where we consider a different  edge $r_G(\sigma_i)$ leading to the root for each such substitution $\sigma_i$, and discarding only the integrity constraints that do not produce useful residues for the query.

\subsubsection{Completeness}
\label{sec:completeness}

\begin{definition}\label{def:completeness}
Let $D =\langle EDB, IDB, IC \rangle$ be a database and let $b$ be the belief logic program associated with $D$ and extended with the integrity constraints in $IC$.

For a query $Q: q(\vec{z}) \der F(\vec{y})$ and its rewriting $Q': q'(\vec{z}) \der G(\vec{y})$, let $\vec{x}$ be an answer to $Q$. The completeness of the set of answers retrieved by executing $Q'$ is measured by the belief by $b$ in $q'(\vec{x})$.
\end{definition}

As usual, for the query $Q'$ we define a fictitious rule 
$
r'_G: \ [1,1] $ $q'(\vec{z}) \der G(\vec{y})
$.
Intuitively, the completeness of the set of answers returned after executing the rewritten queries obtained from $Q$ is measured by the ``likelihood'' that an answer to the original query $Q$ is also an answer to the rewritten query and is thus retrieved by processing it.

\begin{description}
\item [Removal of an atom] 
Dually to the analysis of correctness, if $Q'$ has been obtained from $Q$ only by removing atoms, then $belief(q'(\vec{x})) = 1$.
\item [Introduction of an atom]
Assume that restrictions and/or (extensional) atoms have been introduced in the query.
Given a generic answer $\vec{x}$ to $Q$, to compute the value of belief in $q(\vec{x})$, we consider again a proof DAG having as root $q'(\vec{x})$. The graph has, at the first level of a-nodes, the literals in the body of $Q'$; every atom that has been introduced in the rewritten query is the root of a subtree corresponding to the integrity constraints from which the atom has been obtained as a residue.

A lower bound on the value of the belief is computed on the graph similarly as described for the evaluation of correctness of a rewriting that removes an atom.

Notice that the DAG does not include the r-node and corresponding edge for the extensional atoms that have been inserted. In the presence of these facts, the lowest bound on the value of completeness corresponds again to the case where not all the atoms that have been inserted are true, and thus is equal to 0. Dually to correctness, to compute a meaningful value, we determine the lower bound by considering the belief in $q(\vec{a})$ by the belief logic program extended with the integrity constraints and without the facts denoting the inserted predicates.

\begin{example}\label{ex:removalCompleteness}
For instance, for the query $Q_1$ of Example~\ref{ex:dbBookUn} rewritten as $Q'_1: q'_1(x_1, x_2, y_2, y_3) \der PCMember(x_1, x_2), Conference(x_2, y_2, y_3), Authoritative(x_1, $ $y_2), y_3 = Australia$ by introducing the atom $Authoritative(x_1, y_2)$, we build a corresponding graph, depicted in Figure~\ref{fig:DAGRemovalComp}, and we compute a lower bound on the belief equal to 0.8.\hfil\qed

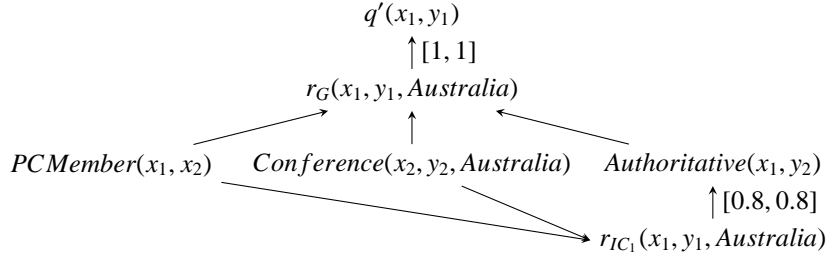
\begin{figure}[h!]
\centering
\begin{tikzpicture}[->, >=stealth,auto,%
    node distance=1cm,%
    every stat/.style={draw=none,text=black, minimum size, inner sep=0pt},%
    every edge/.style={draw,solid}%
    ]
    
  \node         (Pc) []  {$PCMember(x_1, x_2)$};
  \node         (C) [right of= Pc, xshift=3cm]     {$Conference(x_2, y_2, Australia)$};       
  \node         (At) [right of= C, xshift=3cm] {$Authoritative(x_1,y_2)$};

  \node         (rg) [above of= C]     {$r_G(x_1, y_1, Australia)$};
  \node         (q) [above of= rg]     {$q'(x_1, y_1)$};  
  \node         (ic1) [below of= At]     {$r_{IC_1}(x_1, y_1, Australia)$};  
    
  \path[->] (Pc)  edge[]   (rg)
				  edge	(ic1.west)
			(C)   edge[]   (rg)
				  edge	(ic1.west)
			(At)  edge  (rg)
			(rg)  edge[]  node[right] {$[1,1]$} (q)				  
			(ic1) edge[] node[right] {$[0.8,0.8]$}  (At)
;   
\end{tikzpicture}
\caption{Proof DAG to compute a lower bound on the belief in Example~\ref{ex:removalCompleteness}.\label{fig:DAGRemovalComp}}
\end{figure}
\end{example} 

Notice also that if a restriction $A = x \theta u$ is introduced in the query, its combined belief is computed taking into account all the residues having as head a restriction on the same variable that implies $A$.
For instance, for the query $Q_3$ we can derive $y_2 \geq 30$ as a residue from both $IC_4$ and $IC_5$.

\item [General rewriting] 
Computing the belief is dual to the case of correctness.

If the rewritten query $Q': q'(\vec{z})\der G(\vec{y})$ has been obtained from $Q$ without introducing atoms, then each of its answers $\vec{x}$ is valid also for the query $Q'$ and $belief(q'(\vec{x})) = 1$.

Otherwise, at least an atom has been introduced in $Q'$.

Let $\vec{x}$ be a generic answer to $Q$ and  let $\sigma$ be any substitution for the variables of $Q$ with $dom(\sigma) = \vec{y}$ and $\vec{z} \sigma = \vec{x}$, such that $\models  F(\vec{x}) \sigma$. 

To compute a lower bound on the belief in $q'(\vec{x})$ we build as usual a proof DAG, and set as false all the formulae corresponding to bodies of  constraints for which it is not explicitly known a combined belief.

\begin{enumerate}
\item The graph has root $q'(\vec{x})$, reached by an edge labeled $[1,1]$ from an r-node $r'_G$
\item The first level of a-nodes includes the atoms in the body of $Q'$
\item \label{item:completenessGeneral} For each restriction/extensional atom that has been introduced in $Q'$, 
\begin{enumerate} 
\item \label{item:completenessGeneral1} include in the proof DAG the subtrees corresponding to the integrity constraints used to derive it.
For every atom in the body of these constraints:
\begin{enumerate}
\item either it occurs in the body of $Q$ (and thus holds true with certainty) 
\item or it is a restriction/extensional atom introduced in the body of $Q'$, and it is handled at step~\ref{item:completenessGeneral}.
\end{enumerate}
\end{enumerate}
\end{enumerate}
\end{description}

\subsubsection{Semantic query transformation}
\label{sec:queryTransformation}

We describe in Algorithm~\ref{algo:GreedySQT} the pseudo-code of the function that selects a rewriting for a user's query $Q$ by exploiting the available residues.
The inputs to Algorithm~\ref{algo:GreedySQT} are the query $Q$, a threshold  on correctness $T_{corr}$ and  on completeness $T_{comp}$ chosen by the user for the rewriting, the set $R_{all}$ of residues associated to the query by the phase of semantic compilation. We assume that each residue is associated with the set of integrity constraints it has been derived from and with the substitution of variables used to apply partial subsumption between the query and the constraints -- so that we can identify the atoms of the query that derived the residue.
The algorithm returns the transformed query $Q'$ along with the estimates on correctness and completeness of the applied rewriting.

The algorithm is greedy and chooses the residues to apply according to an evaluation of cost based on information on the dataset, regarding the cardinality of the extensions of the predicates and the availability of indexes, and/or possibly heuristics of the query processor.

First it applies the residues that eliminate predicates from the query, choosing them by decreasing order of extension, i.e., from the most cost-effective to the least promising one. Then it introduces restrictions and subsequently extensional atoms.
For each possible basic transformation it computes a lower bound on the correctness (completeness) of a generic answer (set of answers) that could be returned by the rewritten query and proceeds until a threshold set by the user on correctness and completeness is not exceeded.

The auxiliary function in Algorithm~\ref{algo:InitializeDAG}, called at line~\ref{callInitialize} of Algorithm~\ref{algo:GreedySQT}, builds an initial proof DAG $G$ that is used to compute the values of correctness and completeness associated with a rewriting. We consider a single DAG to compute both measures. The DAG has two roots $q$ and $q'$, reached respectively by an edge from a r-node $r_g$ and $r'_g$ having as children the literals in the body of the query $Q$.  To each a-node $i$ of the graph we associate a set $I_i$, initially empty, of belief intervals of integrity constraints that generate residues that derive it.

Functions in algorithms~\ref{algo:ValidRemoval} and~\ref{algo:ValidInsertion} are invoked when an atom is removed from or, respectively, inserted in the query. They update the proof DAG $G$ as described in Section~\ref{sec:correctness} and~\ref{sec:completeness} and check whether the removal or insertion of an atom is valid. For simplicity, in the algorithms $T_{corr}, T_{comp}, G$ are assumed to be global variables. 

Notice that in the case of a database with an empty IDB, it is possible to compute the lower bound on the belief without actually building the DAG. The construction of the DAG is instead useful to handle the general case of a database where intensional predicates can occur in the rules and constraints, which will be described in the following Section~\ref{sec:approximateQAExtensions}.

Notice also that in Algorithm~\ref{algo:GreedySQT}, once an atom is introduced in the query, it can trigger the application of new residues, whose body becomes empty if it unifies with the new inserted atom. Thus, the cycle of application of transformations can be iteratively executed until there is no new residue that can be applied without exceeding the thresholds.

\begin{remark}
The procedure for the choice of a rewriting is not fully optimized.
In particular, a more effective approach consists in computing in the phase of semantic compilation (when the query has not been yet submitted) all the atoms that can be derived by applying a given residue, relieving the need, in the current algorithm, to iterate the check for the existence of new residues with empty body during the phase of semantic transformation. That is, for each database predicate we can consider both the residues $R$ directly derived for it as described in Section~\ref{sec:SQO} and, recursively, the residues associated with the head of $R$, and the residues associated with the head of these ones and so on. To each predicate we can associate a tree of derivations, where at each step of the chain a new atom is introduced. The last atom of the chain can be possibly chosen for a rewriting that removes or introduces the atom. We need to modify Algorithm~\ref{algo:GreedySQT} accordingly, so that a sequence of actions instead of a single action is evaluated, both in the initial sorting phase based on the estimation of the cost of processing a query, and in the following phase of validation of the transformation of removal and/or insertion of atoms defined by the sequence.
\end{remark}

\begin{algorithm}
\caption{GreedySemanticQueryTransformation}\label{algo:GreedySQT}

\begin{algorithmic}[1]

\Require {user's query $Q$; threshold on minimum value of correctness $T_{corr}$; threshold on minimum value of completeness $T_{comp}$; set $R_{all}$ of residues associated to the query}
\Ensure{rewritten query $Q'$; correctness $corr$; completeness $comp$ of the rewriting}

	\Statex
	\Procedure{GreedySemanticQueryTransformation}{$Q, T_{corr}, T_{comp}, R_{all}$}
	\State $Q' \der Q$
	\State $R \der$ set of the residues in $R$ with empty body
	\State $R_{new} \der R_{all} \setminus R$
	\State $corr \der comp \der 1$
	
	\State $G \der$ \Call{InitializeDAG}{$Q$} \label{callInitialize}
	\State $Dep_{corr} \der \emptyset$
	
	\While {$R \neq \emptyset$}
	
		\LongState{$D \der$ residues in $R$ whose head is an atom occurring
		 in the body of $Q$ and that has a large extension in the database}
		\State Sort $D$ in decreasing order of the value of the extension in the head
		\While{$D\neq \emptyset$} 
			\State $A \der D.pop()$
			\State $valid, corr \der$ \Call{ValidRemoval}{$A, corr$}
			\If{$valid$}
				\State $Q' \der Q' \setminus A$
			\EndIf	
		\EndWhile

		\LongState{$A_r \der$ residues in $R$ whose head is a restriction on an attribute of predicates in $Q'$ for which an index is available}
		\While{$A_r\neq \emptyset$} 
			\State $A \der A_r.pop()$
			\State $valid, corr, comp \der$ \Call{ValidInsertion}{$A, corr, comp$} 
			\If{$valid$}
				\State $Q' \der Q' \cup A.head$
				\LongState{$R_{new} \der$ set of residues obtained by applying
								 partial subsumption between each residue in $R_{new}$ and $A$}
			\EndIf	
		\EndWhile
		
		\LongState{$A_p \der$ residues in $R$ whose head is an atom not occurring in $Q$, with small extension and possibly indexes on attributes}
		\State Sort $A_p$ in increasing order of the value of the extension in the head	
		\While{$A_p\neq$} 
			\State $A \der A_p.pop()$
			\State $valid, corr, comp \der$ \Call{ValidInsertion}{$A, corr, comp$} 
			\If{$valid$}
				\State	$Q' \der Q' \cup A.head$
				\LongState{$R_{new} \der$ set of residues obtained by applying
		 partial subsumption between each residue in $R_{new}$ and $A$}

			\EndIf	
	\EndWhile \Comment{continue}
		
	\algstore{bkbreak}	
	\end{algorithmic}
	\end{algorithm}

	\begin{algorithm}[h]
	\begin{algorithmic}[1]

	\algrestore{bkbreak}
		\State $R_{temp} \der$ residues in $R_{new}$ with empty body\ \Comment{We attach to each residue}
		\LineComment {also the annotation of a residue in $R$ -- if any -- that has not been used in }
		\LineComment{the rewriting and has the same head}
		\LongState{$R \der R_{temp}$  $\cup$ \{residues in $R_{all} $ for which no index was available for the predicates occurring in $Q'$\}}
		
 	\EndWhile	
	\State \Return $Q', corr, comp$
	
\EndProcedure

\end{algorithmic}

\end{algorithm}

\begin{algorithm}[h]
\caption{InitializeDAG}\label{algo:InitializeDAG}

\begin{algorithmic}[1]

\Require {user's query $Q$}
\Ensure{Proof DAG}

	\Statex
	\Procedure{InitializeDAG}{$Q$}

	\State Create two a-nodes $q$ and $q'$ in $G$
	\State create an r-node $r_g$ with an edge labeled $[1,1]$ towards $q$
	\State create an r-node $r'_g$ with an edge labeled $[1,1]$ towards $q'$	
	\For{each literal $L_i$ in the body of $Q$}
		\LongState{create an a-node $i$ and associate with it a set of belief intervals $I_i$, initially empty}
		\State create an edge from $i$ to $r_g$
		\State create an edge from $i$ to $r'_g$
	\EndFor
	
	\State \Return $G$	
	\EndProcedure
\end{algorithmic}
\end{algorithm}

\begin{algorithm}
\caption{ValidRemoval}\label{algo:ValidRemoval}

\begin{algorithmic}[1]

\Require {residue $A$ whose head can be removed from the query, current value of correctness $corr$}
\Ensure{Return $0$ if the thresholds are exceeded when the atom is removed; otherwise return $1$. Return also the new value of correctness $corr$.}

	\Statex
	\Procedure{ValidRemoval}{$A, corr$}

	\State $H \der A.head$
	\State $B_H \der$ set of belief intervals associated to $A$
	
	\LineComment{The a-node $H$ is already in $G$ with an associated set $I_H$}
	
	\If{$\nexists$ an edge between $H$ and $r'_g$} \Comment $i.e., I_H \neq \emptyset$
		\LineComment{a-node $H$ has been already removed; thus, we update its combined belief}
		\State $corr_{new} \der corr \cdot \frac{\phi_V(I_H \cup B_H)}{\phi_V(I_H)}$ 
		\If{$corr_{new} < T_{corr}$}
			\State \Return $0, corr$
		\EndIf
		\State $I_{H} \der I_H \cup B_H$
		\State \Return $1, corr_{new}$ 			
	\Else
		\State $corr_{new} \der corr \cdot \phi_V(B_H)$ 
		\If{$corr_{new} < T_{corr}$}
			\State \Return $0, corr$
		\EndIf
		\State remove the edge between $H$ and $r'_g$
		\State $I_H \der B_H$
		\State \Return $1, corr_{new}$		
		
	\EndIf
			
	\EndProcedure
\end{algorithmic}
\end{algorithm}

\begin{algorithm}
\caption{ValidInsertion}\label{algo:ValidInsertion}

\begin{algorithmic}[1]

\Require {residue $A$ whose head can be inserted in the query, current value of correctness $corr$ and completeness $comp$ }
\Ensure{Return $0$ if the thresholds are exceeded when the atom is inserted; otherwise return $1$. Return also new value of $corr$ and $comp$}

	\Statex
	\Procedure{ValidInsertion}{$A, corr, comp$}

	\State $H \der A.head$
	\State $B_H \der$ set of belief intervals associated to $A$
	\If{the a-node $H$ is not in $G$}
		\State add a-node $H$ to $G$ with $I_H = \emptyset$
	\EndIf
	
	\If{$\exists$ an edge between $H$ and $r'_g$} 
			\LineComment{a-node $H$ has been already inserted; thus, we update its combined belief}
			\State $comp_{new} \der comp \cdot \frac{\phi_V(I_H \cup B_H)}{\phi_V(I_H)}$ 
			\If{$comp_{new} < T_{comp}$}
				\State \Return $0, corr, comp$
			\EndIf
			\State $I_{H} \der I_H \cup B_H$
			\State \Return $1, corr_{new}, comp_{new}$ 				
	\Else
		\State $comp_{new} \der comp \cdot \phi_V(I_H \cup B_H)$ 
		\If{$comp_{new} < T_{comp}$}
			\State \Return $0, corr, comp$
		\EndIf	
		\State add an edge between $H$ and $r'_g$
		\State $I_H \der B_H$
		\State \Return $1, corr_{new}, comp_{new}$
		
	\EndIf

	\EndProcedure
\end{algorithmic}
\end{algorithm}

\section{Extension of approximate query answering}
\label{sec:approximateQAExtensions}

In this section we extend the approach for approximate query answering to the case that the IDB of the database is not empty:  we first  require that no negated intensional predicate occurs in the rules and integrity constraints (Section~\ref{sec:extensionIDB}), and we then present some hints at how to deal  with some issues emerging  in the presence of negated intensional predicates (Section~\ref{sec:extensionNegInt}).

\subsection{Extension to a non empty IDB}
\label{sec:extensionIDB}

In the following, we assume that the IDB of the database is not empty and contains rules with belief interval $[1,1]$.  Also, the rules in the IDB, the integrity constraints and the queries cannot include negated  intensional predicates.

As in the simpler case of a database consisting only of the EDB and IC, we consider the two standard phases of SQO, semantic compilation and semantic transformation, and we show how to adapt them to handle the presence of uncertain semantic knowledge.

\subsubsection{Semantic Compilation}
\label{sec:extensionsIDBSC}
The first phase of SQO associates residues to predicates and annotates them with
the name (and belief interval) of the integrity constraint from which they have been obtained.
In the classical phase of compilation the integrity constraints are preprocessed so that the intensional predicates occurring in their definition are elaborated into solely extensional predicates. 
Differently from the standard procedure of semantic compilation, however, in the presence of uncertain semantic knowledge the expansion cannot be directly performed, propagating the belief intervals, 
otherwise inconsistent information on the database can be inferred, as the following example shows.

\begin{example}
The Italian government is arranging a spending review to reduce current public expenditure. 
The data on the budget expenditure of the government departments is collected in a database having, among others, the extensional predicates $Outcome/2$, $CivilDefense/2$, $Renewal/3$, $MilitaryVehicle/2$, and an intensional predicate $DefenseDepartmentCost/2$ (abbreviated $DDC/2$) defined by the two rules (holding true with certainty):
\[
\begin{array}{ccll}
r_{A_1}: &[1,1] & DDC(code, cost) \der & Outcome(code, cost), CivilDefense(code, city)\\
r_{A_2}: &[1,1] & DDC(code, cost) \der & Outcome(code, cost), Renewal(code, sector,\\
		 &		& 					   &  vehicleType), MilitaryVehicle(type, cost)\\
\end{array}
\]
To examine the current trends of expenses, the government technicians mine the database, finding the following (uncertain)  integrity constraint.
\[
\begin{array}{ccl}
r_{IC_1}: &[0.8,0.8] & cost < 10K \der DDC(code, cost)\\
\end{array}
\]
However, the high degree of certainty of the constraint derives from the fact that the department's budget includes several low-cost items for civil defense, but a few huge outlays for renewal of military equipment. 
Thus, it is not  correct to rewrite $r_{IC_1}$ into the two following annotated rules:
\[
\begin{array}{ccll}
r_{IC_{11}}: &[0.8,0.8] & cost < 10K \der & Outcome(code, cost), CivilDefense(code, city)\\
r_{IC_{12}}: &[0.8,0.8] & cost < 10K \der & Outcome(code, cost), Renewal(code, sector,\\
		 &		& 					   &  vehicleType), MilitaryVehicle(type, cost)\\
\end{array}
\]
even under the assumption that we consider belief combination functions $\phi$ such that $\phi([v,w],[v,w]) = [v,w]$, otherwise one might erroneously deduce that current Italian expenses for orders of new military equipment (planes, \ldots) is not actually exceedingly high.

In fact, if the integrity constraint has been mined from a database, the rewriting is not reasonable unless either the association rule has been mined with high support and confidence and the atoms in the extension of the predicate $DDC$ are uniformly distributed among the extensions $E_{1}$ and $E_{2}$ of the body of its defining rules in the database, or there is a priori available information on the fact that the atom $cost < 10K$ holds true with comparable likelihood for elements in both $E_{1}$ and $E_{2}$.\hfill\qed
\end{example}

Thus, semantic compilation computes the residues as integrity constraint fragments by applying partial subsumption between every integrity constraint and either the body of a rule, or the body of dummy rules $r \der r$ where $r$ is an extensional predicate or an intensional predicate (as it is not reasonable to expand the intensional predicates occurring in the integrity constraints) or the body of dummy rules $\neg r \der \neg r$ where $r$ is an extensional predicate. .
All the residues are annotated with the name of the integrity constraint they are derived from.

\subsubsection{Semantic transformation}
\label{sec:extensionsIDBST}
The second phase of SQO, semantic transformation, rewrites a user's query by using the residues associated with its predicates. 

In the classical approach all the intensional predicates occurring in the body of the original query are expanded, generating  a set of queries consisting only of extensional predicates, such that the union of the answers to the queries of this set represents the set of answers of the original one. 
The queries of the set are then transformed by adding or deleting predicates from their body on the basis of the residues associated with them.

With respect to the standard technique, in the presence of uncertain semantic knowledge a few issues have to be taken care of.

The computation of the residues and the choice of which of them  to apply to the query is not necessarily postponed until it has been completely elaborated into a set of queries consisting of extensional predicates, since, if we cannot unfold the intensional predicates in the integrity constraints, they can occur also in the body or head of the possible residues associated with the predicates of the query. 
Thus, whenever an intensional predicate occurs in the body of the query, we compute the possible residues whose body is subsumed by the predicate. 
Furthermore, since we admit integrity constraints with an intensional predicate as the head of the clause, if a residue whose body subsumes the body of the query has an intensional atom in the head that can be unified with an atom occurring in the query, then it can be removed from the query's body before it is further expanded.

\begin{example}\label{ex:dbBookIDB}
Consider the bibliography database of Example~\ref{ex:dbBook} and assume that its IDB includes the following rules.

\noindent IDB:
\begin{align*}
r_{A_1} : \ [1,1] \  Coauthors(x_1,y_1) \der & Author(x_1, x_2, x_3), Author(y_1, y_2, y_3), AuthorPaper(x_1, z_2) ,\\
\notag & AuthorPaper(y_1, z_2) 
\end{align*}
``Two authors that wrote a paper together are coauthors.''

\begin{align*}
r_{A_2} : \ [1,1] \  WorkWellTogether(x_1,y_1) \der & Authoritative(x_1, x_2), Authoritative(y_1, x_2)
\end{align*}
``Two authoritative authors on the same subject work well together.''

\begin{align*}
r_{A_3} : \ [1,1] \  WorkWellTogether(x_1,y_1) \der & AuthorPaper(x_1, x_2), AuthorPaper(y_1, x_2),\\
\notag & Coauthors(x_1, y_1), Bestseller(x_2)
\end{align*}
``Two coauthors that wrote together a bestseller work well together.''\\

\noindent and assume that there are also the following integrity constraints:
\begin{align*}
r_{IC_7} : \ [0.7,0.7] \  Bestseller(z_2) \der & Coauthors(x_1,x_2), Authoritative(x_1, y_2),\\
\notag & AuthorPaper(x_2, z_2) , Paper(z_2, y_2) , y_2 = math
\end{align*}
``A coauthor of an  author that is authoritative in Math writes Math papers that are bestseller.''

\begin{align*}
r_{IC_8} : \ [0.8,0.8] \  Coauthors(x_1,y_1) \der & Author(x_1, x_2, x_3), Author(y_1, y_2, y_3),  AuthorPaper(x_1, z_2), \\
\notag & Paper(z_2, s) , AuthorPaper(y_1, z_3) , Paper(z_3, s)
\end{align*}
``Two authors that write papers on the same subject are coauthors.''

Consider the following queries:

\noindent $Q_5$: ``Find  coauthors authoritative in Math that wrote together a paper on Math."
\begin{align}\label{eq:Q5}
Q_5:  \der &  Coauthors(x_1,x_2), Authoritative(x_1, y_2), Authoritative(x_2, y_2),\\
 \notag &   AuthorPaper(x_1, z_2) , AuthorPaper(x_2, z_2) , Paper(z_2, y_2), y_2 = math
\end{align}

Notice that query $Q_5$ has, among others, an associated residue $Bestseller(z_2) \der   Authoritative(x_1, y_2), AuthorPaper(x_2, z_2) , Paper(z_2, y_2) , y_2 = math$ from the integrity constraint $r_{IC_7}$ having the intensional predicate $Coauthors(x_1,x_2)$ in its body, and the residue might be exploited to optimize the query by introducing the atom $Bestseller(z_2)$.

\noindent $Q_6$: ``Find  coauthors that wrote a paper on TCS."
\begin{align}\label{eq:Q6}
Q_6:  \der &  Coauthors(x_1,y_1), Author(x_1, x_2, x_3), Author(y_1, y_2, y_3),\\
\notag &    AuthorPaper(x_1, z_2), Paper(z_2, s_2), AuthorPaper(y_1, z_3) ,\\
\notag &   Paper(z_3, s_2), s_2 = TCS
\end{align}

Query $Q_6$ has, among others, an associated residue $Coauthors(x_1,y_1) \der Author $ $(x_1, x_2, x_3), Author(y_1, y_2, y_3), $ $ AuthorPaper(x_1, z_2), Paper(z_2, s_2) , AuthorPaper$ $(y_1, z_3) , $ $Paper(z_3, s_2)$ from the integrity constraint $r_{IC_8}$, which can be exploited to simplify the execution of the query by removing the intensional predicate $Coauthor$ before it is further expanded.
\hfill\qed
\end{example}

\subsubsection*{Expansion of a query}
\label{sec:extensionsSTexpansion}

The steps of the procedure of semantic transformation that expands the query and associates with it a set of residues   are the following.
\begin{enumerate}
\item Let $Q: \der L_1, \ldots, L_m$ be an input query. Initialize a set $R$ of removed predicates to the empty set.
\item \label{item:extensionsIDBExpansion1} Compute the residues associated with the predicates in the body of the query. 
\item \label{item:extensionsIDBExpansion2} For each predicate $L_j$ ($1 \leq j \leq m$) that is an intensional predicate such that there is a residue (not necessarily associated with $L_j$) whose body subsumes $R \land \bigwedge_{i=1}^{m} L_i$ and has $L_j$ as head 
\begin{enumerate}
\item check whether the predicate can be removed from the body of the query, and if this is the case, rewrite the query by eliminating this atom and add the atom to set $R$. 
Notice that the choice whether to remove or not the predicate can be performed greedily, on the basis of an analysis of correctness and completeness determined by a threshold set by the user (shown next), or all the possible alternatives of rewriting can be exhaustively examined.
\end{enumerate}
\item If the body of the query still contains intensional predicates 
\begin{enumerate}
\item expand them according to their defining rules, and add them to set $R$. 
\item Remove disjunctions in the query, obtaining a set of queries expressed as conjunctions of extensional and intensional predicates (generated by the expansion). Repeat step~\ref{item:extensionsIDBExpansion1} for each query of this set (each one is associated with a copy of set $R$).
\end{enumerate}

\end{enumerate}

Similarly to the analysis for an empty IDB, to determine whether a removal is valid or not (after possibly a sequence of removals and/or expansions of intensional predicates in the query has been already performed), we evaluate the following estimates of correctness and completeness, as in Section~\ref{sec:SQO-combine}.

Let $Q: q \der L_1, \ldots, L_m$ be the input query to the procedure of expansion and let $Q'$ be a query examined at step~\ref{item:extensionsIDBExpansion2} of the procedure of expansion after a candidate intensional predicate has been removed and a (possibly empty) set of intensional predicates has been already expanded or removed.

\subsubsection*{Correctness for a removal of an intensional predicate}
\label{sec:correctnessRemovalInt}
Given an answer $\vec{x}$ to $Q'$, we estimate the correctness of the rewriting by computing the belief in $q(\vec{x})$ by the blp corresponding to the database extended with the uncertain integrity constraints. 

To this end, we build a proof DAG:
\begin{enumerate}
\item The graph has root $q(\vec{x})$, reached by an edge labeled $[1,1]$ from an r-node $r_G$
\item The first level of a-nodes includes the atoms in the body of $Q$.

 Clearly, each restriction/extensional atom in $Q$ is present also in the body of $Q'$ and $\vec{x}$ satisfies it; also, $\vec{x}$  satisfies the intensional predicates in $Q$ that have not been removed in $Q'$.
\item \label{item:extensionsCorrectness1} For each intensional predicate of $Q$ that does not occur in $Q'$
\begin{enumerate} 
\item \label{item:extensionsCorrectness2} if it has been already removed  by SQO, include in the proof DAG the subtrees corresponding to the integrity constraints used to derive it. They have as leaves either extensional predicates/restrictions or intensional predicates in $Q'$ (which $\vec{x}$ satisfies) or intensional predicates not occurring in the body of $Q'$ and that are handled as in step~\ref{item:extensionsCorrectness1}..
\item if it has been expanded, build the chain of expansions that led from $Q$ to $Q'$, where at each level of the tree of the expansion an atom can be, again, an extensional predicate/restriction or an intensional predicate occurring in the body of $Q'$ (and holds true for $\vec{x}$), or an intensional predicate not occurring in the body of $Q'$ and it is handled as in step~\ref{item:extensionsCorrectness1}.
\end{enumerate}
\end{enumerate}

As usual, we compute the belief in $q(\vec{x})$ on this DAG: given the assumptions on the belief combination function and the fact that the rules are annotated by belief interval $[1,1]$, it yields a lower bound for the belief in this atom by the blp.

\begin{example}
\label{ex:extensionsCorrectnessRemoval}

Consider the following query:

\noindent $Q_7$: ``Find  authors that wrote a paper on Logic Programming (not necessarily together) and work well together."
\begin{align}\label{eq:Q7}
Q_7:  \der &  WorkWellTogether(x_1,y_1), Author(x_1, x_2, x_3),\\
\notag &   Author(y_1, y_2, y_3), AuthorPaper(x_1, z_2), Paper(z_2, s_2) ,\\
\notag &  AuthorPaper(y_1, z_3) , Paper(z_3, s_2), s_2 = logicProgramming
\end{align}

Semantic transformation applies the steps of the aforementioned procedure to expand the query into a set of queries expressed as conjunctions of extensional predicates.

First, it computes the residues associated to the predicates in the body of the query.
Since the predicates of the query have no residue with the intensional predicate $Work$ $WellTogether$ as head, the intensional predicate is expanded, yielding two queries, $Q_{71}$ and $Q_{72}$:
\begin{align*}
Q_{71}:  \der &  Authoritative(x_1, t_2), Authoritative(y_1, t_2), Author(x_1, x_2, x_3),\\
\notag &   Author(y_1, y_2, y_3), AuthorPaper(x_1, z_2), Paper(z_2, s_2) ,\\
\notag &  AuthorPaper(y_1, z_3) , Paper(z_3, s_2), s_2 = logicProgramming
\end{align*}
\begin{align*}
Q_{72}:  \der &   AuthorPaper(x_1, t_2), AuthorPaper(y_1, t_2), Coauthors(x_1, y_1),\\
\notag &  Bestseller(t_2), Author(x_1, x_2, x_3), Author(y_1, y_2, y_3),\\
\notag &    AuthorPaper(x_1, z_2), Paper(z_2, s_2) , AuthorPaper(y_1, z_3) ,\\
\notag &  Paper(z_3, s_2), s_2 = logicProgramming
\end{align*}

As regards query $Q_{71}$, the procedure computes the residues associated with the new generated predicates. As only extensional predicates occur, the procedure stops.

Concerning query $Q_{72}$, the residue $Coauthors(x_1,y_1) \der$ derived from the constraint $r_{IC_8}$ can be exploited to remove the atom. To check whether the removal is valid, we build the  proof DAG in Figure~\ref{fig:ExtensionsDAGRemovalCorr}, and we check whether the computed lower bound on the belief, $0.8$, does not exceed a threshold on correctness set by the user.
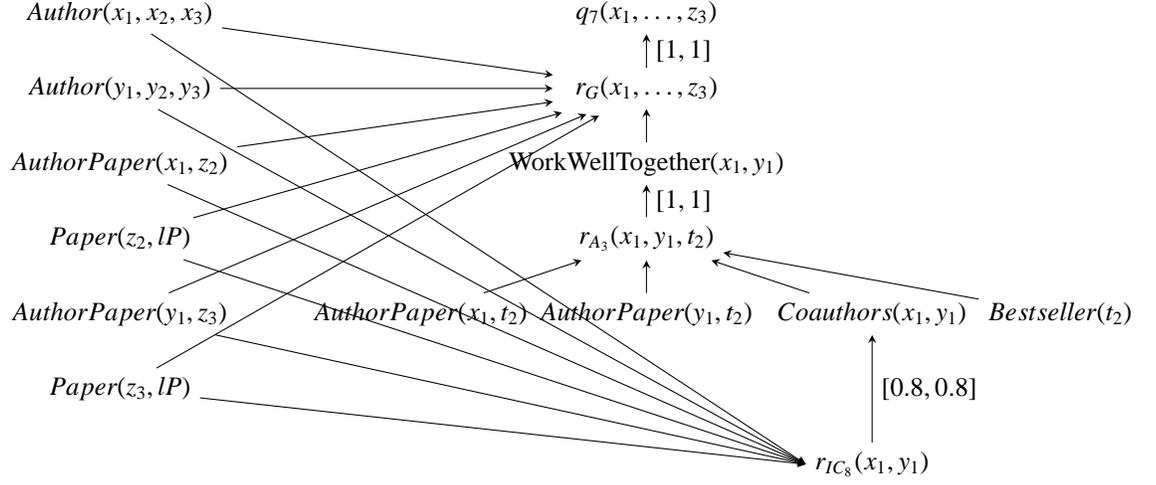
\begin{figure}[h!]
\begin{tikzpicture}[->, >=stealth,auto,%
    node distance=1cm,%
    every stat/.style={draw=none,text=black, minimum size, inner sep=0pt},%
    every edge/.style={draw,solid}%
    ]
    
  \node         (A1)                    {$Author(x_1, x_2, x_3)$};
  \node         (A2) [below of= A1]  {$Author(y_1, y_2, y_3)$};
  \node         (Ap1) [below of= A2] {$AuthorPaper(x_1, z_2)$};
  \node         (P1) [below of= Ap1] {$Paper(z_2, lP)$};
  \node         (Ap2) [below of= P1] {$AuthorPaper(y_1, z_3)$};
  \node         (P2) [below of= Ap2] {$Paper(z_3, lP)$};  

  \node         (W) [right of= Ap1, xshift=6cm]     {WorkWellTogether($x_1, y_1$)};  
  \node         (rg) [above of= W]     {$r_G(x_1, \ldots, z_3)$};
  \node         (q) [above of= rg]     {$q_7(x_1, \ldots, z_3)$};  

  \node         (ra3) [below of= W]     {$r_{A_3}(x_1, y_1, t_2)$};

  \node         (Ap4) [below of= ra3] {$AuthorPaper(y_1, t_2)$};
  \node         (Ap3) [left of= Ap4, xshift=-2cm] {$AuthorPaper(x_1, t_2)$}; 
  \node         (C) [right of= Ap4, xshift=2cm] {$Coauthors(x_1, y_1)$};  
  \node         (B) [right of= C , xshift=1.5cm] {$Bestseller(t_2)$};  

  \node         (ic8) [below of= C, yshift=-1cm]     {$r_{IC_8}(x_1, y_1)$};
     
  \path[->] (A1)  edge[shorten >= 5pt]   (rg)
				  edge	(ic8.west)
			(A2)  edge[shorten >= 5pt]   (rg)
				  edge	(ic8.west)					  					
			(Ap1) edge[shorten >= 5pt]   (rg)
				  edge	(ic8.west)
			(P1)  edge[shorten >= 5pt]   (rg)
				  edge	(ic8.west)
			(Ap2) edge[shorten >= 5pt]   (rg)
				  edge	(ic8.west)
			(P2)  edge[shorten >= 5pt]   (rg)
				  edge	(ic8.west)	
			(W)   edge[]   (rg)			  				  				  							  
			(Ap3) edge	(ra3)
			(C)   edge	(ra3)
			(Ap4) edge	(ra3)
			(B)   edge	(ra3)				  				  				  							  

			(rg)   edge[]  node[right] {$[1,1]$} (q)				  
			(ra3)  edge[] node[right] {$[1,1]$}  (W)		
			(ic8)  edge[] node[right] {$[0.8,0.8]$} (C)
;   
\end{tikzpicture}
\caption{Proof DAG to compute a lower bound on the belief of $q_{7}(\vec{x})$ in Example~\ref{ex:extensionsCorrectnessRemoval}.\label{fig:ExtensionsDAGRemovalCorr}}
\end{figure}
\hfill\qed
\end{example}

\subsubsection*{Completeness for a removal of an intensional predicate}
\label{sec:completenessRemovalInt}
The removal of the atom does not impact on completeness: an answer to $Q$ need not be an answer to $Q'$, if it does not correspond to the expansion of intensional predicates leading to $Q'$, but the rewriting has no effect on this.

\medskip

Now, for an input query $Q: q(\vec{z}) \der F(\vec{y})$, let $S = \bigcup_{i=1}^{n} Q_i$ be the set resulting from the expansion, i.e., each $Q_i$ is a conjunction of atoms obtained as described in the procedure reported before.
Let $I_i$ ($1 \leq i \leq n$) be the sets (possibly empty) of atoms removed from each query $Q_i$ during the expansion by exploiting the available residues.

Once a set of queries is obtained, the purpose of SQO is to rewrite them into a form that is easier to process (an effective approach could perform global or multiple query optimization, as in~\cite{HsuKnoblock2000,Sellis1988,Gallaire1984}, but here we do not deal with this issue). 

In fact, we optimize separately each query $Q_i$ of the resulting set by applying the available residues, and we evaluate correctness and completeness of a rewriting $Q'_i$ similarly to the case that the IDB is empty.

First, given a  database $D =\langle EDB, IDB, IC \rangle$, let $b$ be the belief logic program associated with $D$ and extended with the integrity constraints in $IC$.

As regards correctness, given an answer $\vec{x}$ to $Q'_i$,  we measure a lower bound on the belief by $b$ in the original query $q(\vec{x})$.
The lower bound is computed analogously to the case of an empty IDB, with the only difference that the chain of intensional predicates that have been expanded to generate the query and the removed intensional predicates must be taken into account (as done in the analysis for the removal of an intensional predicate during the expansion).

As regards completeness, given an answer $\vec{x}$ to $Q_i$,  we measure a lower bound on the belief by $b$ in $q'_i(\vec{x})$.
This case is also analogous to the basic analysis for an empty IDB, as also both $Q_i$ and $Q'_i$ are composed only by extensional predicates. 
The only difference is that a rewriting can  apply  a residue derived from an integrity constraint whose body contains an intensional predicate that was removed or unfolded during the expansion of $Q$. If the residue allows us to insert an atom, then the proof DAG that we use to compute a lower bound on the belief must include also a subtree having this predicate as root: the subtree will include either the integrity constraints that derived the intensional predicate (it it was removed), or (if it was unfolded) the chain of expansions leading to extensional predicates in $Q_i$.

At this point, we can provide a loose evaluation of the quality of the whole set of rewritten queries as follows.
Let $S' =  \bigcup_{i=1}^n Q'_i$ be the set of rewritten queries corresponding to $Q$.

Given an answer $\vec{x}$ to any of $Q'_i$, the correctness of the rewritings is measured by the belief by $b$ in $q(\vec{x})$.
A lower bound on this value is given by the minimum among the value of correctness estimated for the single queries.

Also, given an  answer $\vec{x}$ to $Q$,  the completeness of the set of answers retrieved by executing the queries in $S'$ is measured as
$belief(\bigvee_{i=1}^{n} q'_i(\vec{x}))$.

Clearly, if $Q$ contains intensional predicates and $\vec{x}$ is an answer to $Q$, $\vec{x}$ is an answer to at least one among the queries in $S$.
Since for each query $Q_i$ in this set and its corresponding rewriting $Q'_i$, a lower bound on the belief by $b$ in $q'_i(\vec{x})$ is known, then a lower bound on the $belief(\bigvee_{i=1}^{n} q'_i(\vec{x}))$ is given by the minimum among the values of completeness of the queries.

\subsection{Extension to  rules and integrity constraints with negated intensional predicates}
\label{sec:extensionNegInt}

In the previous sections we assumed that the rules of the database are annotated with a belief interval $[1,1]$. According to the semantics of the theory of BLP (see Section~\ref{sec:BLP}), this assumption implies that it is not possible to infer  from the EDB and IDB of the database the negation of an intensional predicate, with a non null belief.

In fact, for a  predicate $q$ of a belief logic program, the degree of certainty in the falsity of $q$ is computed by Equation~\eqref{eq:modelBLP} as the sum of the support of the truth valuations where $q$ is false, but we can see that all these supports are equal to 0.
Let $P(q)$ be the set of rules or facts having $q$ as head, and for every truth valuation $I$ where $q$ is false, let $P_I(q)$ be the set of rules or facts in $P(q)$ whose body is true in $I$. 
If $P_I(q) = \emptyset$, then the P-support of the truth valuation $I$ is 0.
Otherwise, since the combined belief and disbelief in $q$ are both 1 (assuming a belief function such that $\phi([1,1],[1,1] = [1,1])$), the P-support of every truth valuation $I$ in which $q$ is false is $s_P(q) = 1-1 = 0$. Thus the model of the blp for $\neg q$ is 0.

Thus, to allow also intensional negated predicates to occur in the body of the rules, integrity constraints and queries (and possibly hold true with a positive belief), we admit also rules with belief interval $[v,v]$ different from $[1,1]$, but still with left and right value not less than $0.5$.

The traditional phases of semantic compilation and semantic transformation must be revised to take into account the new assumption on the structure of the rules.

As regards the phase of semantic compilation, the only difference w.r.t the description in the previous section~\ref{sec:extensionIDB}, is that 
the residues are computed as integrity constraint fragments also by applying partial subsumption between every integrity constraint and the body of dummy rules $\neg r \der \neg r$ where $r$ is an  intensional predicate. 

Concerning the second phase of SQO, semantic transformation, we have to deal with some further issues. 

The classical approach, in the absence of uncertain semantic knowledge, expands all the intensional predicates occurring in the body of the original query, yielding  a set of queries with only extensional predicates, such that the union of the answers to the queries of this set equals the set of answers to the original one. 
In general, if the expansion involves also negated intensional predicates whose unfolding leads to negated existentially quantified variables, then the queries in this set may contain also disjunctions of predicates in their body and require that the quantifiers of the variables are explicitly indicated. 
In such cases, simplification by SQO is still possible, although more problematic.

The expansion of a query in the phase of semantic transformation in the presence of uncertain semantic knowledge presents a few differences  with respect to the standard technique and also w.r.t. the description in the previous subsection.
\begin{itemize}
\item Given the convention of BLP to interpret negation as strong negation and given the semantics associated with the annotated rules, the negated intensional predicates should  be expanded into  a disjunction of the bodies of their defining rules (i.e., differently from the classical procedure, it does not consider the conjunction of their negation). 

\item 
The choice whether to eliminate or not positive intensional predicates in the body of the query, by exploiting the available residues, must take into account the fact that also the expansion of the intensional predicates into extensional ones might reduce the value of correctness of the rewriting, since the rules are annotated with belief interval different from $[1,1]$. If, after the removal of the intensional predicates, it is not possible to expand the other intensional predicates in the query without exceeding the threshold on correctness, it is thus necessary to backtrack to the choices in the rewriting that led to the excess. 

A possible way to avoid erroneous choices in the expansion, and to avoid backtracking, consists in the estimation of a lower bound in the reduction of correctness caused by the expansion of an intensional predicate, which can be computed a priori for each of them, before receiving the query, by computing the lowest belief in the predicate for all its possible expansions into extensional predicates.
At each step,  the procedure of expansion of the query can then choose whether to remove  the intensional predicates by computing the resulting measure of correctness by taking  into account also the lower bound in the belief of  the remaining intensional predicates that will be expanded.

A second alternative is to avoid at all the removal of intensional predicates in the query during its expansion, and avoid the cost of selecting the proper predicates to be eliminated. 
\end{itemize}

The evaluation of the quality of a rewriting also has to consider another issue.
\begin{itemize}
\item A problem derives from the fact that if we unfold the negated intensional predicates during the expansion of the query, then, when we compute the measures of correctness and completeness of a rewriting, we cannot always determine a non trivial lower bound in the combined belief in the negated intensional predicate. In fact, if there can be multiple (but we cannot know how many, without querying the database) instantiations of annotated rules having the (positive) predicate as head, they all contribute to increase the combined belief in the positive intensional predicate and, dually, to reduce the belief in its negation.
\end{itemize}

In general, the steps of semantic transformation for the choice of the rewritings can be performed analogously to the presentation in Section~\ref{sec:extensionIDB} if we assume, for simplicity,  that
\begin{enumerate}
\item the negated intensional predicates occurring in the body of the query are not unfolded during the expansion of the query, but are left as-is in the body of the resulting queries and are evaluated at processing time, and
\item we do not remove intensional predicates during the expansion of the query.
\end{enumerate}

This approach to deal with the presence of negated intensional predicates in the body of rules, integrity constraints and queries implies, however, that the rules of the database do not define with certainty their corresponding intensional predicate. An alternative approach, closer to the practice of deductive databases with certain rules or facts, consists in enforcing the typical  convention of negation as failure instead of the assumption of  strong negation for the rules.
First, we require the same assumptions as in~\cite{ChristiansenMartinenghi2006} on the starred dependency graph of the database -- defined in the same work -- i.e., absence of recursion in the database and no negated intensional predicates whose unfolding introduces negated existentially quantified variables.

We keep belief intervals $[1,1]$ for the rules and, for each intensional predicate $p$ with axioms $p(\vec{x}) \der F_1(\vec{y_1}), \ldots, p(\vec{x}) \der F_k(\vec{y_k}) \ (k \geq 1)$, where $F_i$ are conjunctions of predicates, we define a further rule 
$[0,0] : p(\vec{x}) \der  \neg F_1(\vec{y_1}) \land \ldots \land \neg F_k(\vec{y_k})$, which can also be rewritten, by expanding the defining formulae of $p$, as
a set of rules consisting only of conjunctions.

Then, for the evaluation of a query, whenever a negated (resp. positive) intensional predicate occurs, we refer only to the corresponding rules annotated with belief interval $[0,0]$ (resp, $[1,1]$). 
For simplicity, we make the same assumption as before, that the negated intensional predicates occurring in the body of the query are not unfolded during the expansion of the query, but are evaluated at processing time. The steps of semantic transformation are then analogous to those presented in Section~\ref{sec:extensionIDB}, with the difference in the evaluation of the intensional predicates, due to the change in the semantics of negation.  

\section{Future work}
\label{sec:future-work}

In this section we describe some issues that we plan to study as a future work: in particular Section~\ref{sec:experiments} presents the set of experiments that we shall perform to evaluate the effectiveness of our procedure for approximate SQO. Section~\ref{sec:future-work-extensions} hints at some further extensions of our approach to cover more general structures of databases.

\subsection{Experiments}
\label{sec:experiments}

We have not performed yet an experimental evaluation of our approach for query optimization: in this section we describe the set of experiments that we plan to run, as a future work, to validate our approach.

Each experiment consists in the rewriting of a query according to the procedure of approximate SQO described in Section~\ref{sec:extensionIDB}, and in the execution of the original query and the rewritten one on a database. In our experiments we plan to examine the following issues:
\begin{itemize}
\item We evaluate the benefits in performance provided by our approach by comparing the execution time of the original query and the execution time of the transformed one (which includes also the time spent for the rewriting).
\item The rewriting of a query based on the use of uncertain semantic knowledge does not preserve the semantic equivalence w.r.t the original query. 
To evaluate, experimentally, the impact of the proposed approach on the accuracy of the answers returned to the user, we compute precision and recall of the rewritten query.

As described in Section~\ref{sec:SQO-combine}, we estimate the quality of the transformed query in terms of correctness and completeness by computing lower bounds on values of beliefs.
To determine the quality of our estimates of correctness and completeness, we compare them with the actual measures of precision and recall.

Furthermore, we provide the user with an evaluation of correctness of a single answer that has been retrieved by the transformed query.
We estimate the quality of our measure of correctness by determining whether a returned tuple is actually an answer to the original query.
\end{itemize}

\subsubsection{Setup}
\label{sec:experiments-setup}

We plan to experiment with realistic queries and datasets, taken from the TCP-H decision support benchmark.
We refer to queries $Q_3, Q_4, Q_7, Q_9$
and $Q_{10}, Q_{15}$ of the standard, discarding the computation of aggregate values and order by .

In our experiments, we shall consider the following parameters:
\begin{itemize}
\item $s$: size of the database.
\item $c$: minimum degree of certainty of the uncertain integrity constraints. For instance we shall experiment with $c=0.5$, $c=0.7$, $c=0.9$, $c=0.99$.
\item $T_{corr}, T_{comp}$: thresholds on correctness and completeness of the rewriting that the user should set.
\end{itemize}

We shall resort to the TCP-H data-generator script $dbgen$, which creates a database with a prescribed size.

We infer uncertain semantic knowledge about the database by applying a data mining algorithm, which generates association rules, such as $Claudien$~\cite{Raedt2001} (or $Tertius$ \cite{Flach2001}\footnote{$Tertius$, however, assumes a measure of quality of an association rule different from support and confidence, even if these values can still be computed for the inferred association rules and the user can set a frequency threshold on the association rules that are generated.}). The search space of the data mining algorithm is restricted so that it generates only clauses with at most one atom in the head; also, only association rules with confidence not less than $0.5$ are considered.

Notice that we plan to evaluate the accuracy of our estimates on the quality of the rewritten query and its answers, by considering as a parameter for the experiments the minimum confidence of the annotated rules that can be used to optimize the query.
However, not all the association rules inferred by the data mining algorithm are necessarily useful for the query we intend to examine. If we mine the database with different values of minimum confidence $c$, obviously, it is not necessarily the case that we can infer different uncertain integrity constraints, which can be used to optimize the query, for the different chosen ranges for $c$. Hence, we might obtain the same analysis for different values of the parameter.

To evaluate the accuracy as a function of the minimum degree of certainty of the useful constraints, we can simply alter the database by enforcing one (or more) ad-hoc uncertain integrity constraints to hold with a degree of certainty given by  the value of the parameter we want to test. 
Let $Q$ be the test query among $Q_3, Q_4, Q_7, Q_9, Q_{10}, Q_{15}$. First, for instance for a value $c = 0.7$, we  select only the association rules with confidence $>= 0.7$\footnote{Notice that, given a query, we might also define an inductive bias on the mining algorithm, as specified by the $DLAB$ formalism for $Claudien$, by stating some requirements on the predicates occurring in the association rule, so that it induces association rules that are possibly interesting for the query.}. If there is no annotated rule with confidence exactly $0.7$ that is useful for $Q$, we introduce (at least) one ad-hoc uncertain constraint with this confidence.
We might define a restriction on the values of a  variable $y$ of a predicate $p$ in the body of the query (e.g., $y \theta v \der p(\vec{x})$), and modify all the values of the variable in $p$ in the tuples of the database so that a fraction equal to $c$ of the tuples satisfies the restriction; otherwise, we can select two (or more) predicates $p,q$ occurring in the body of the query and we define an integrity constraint involving them (as, e.g., $q(\vec{z}) \der p(\vec{x})$), changing the tuples of the database so that it holds for a fraction $c$ of the tuples.

\subsubsection{Experiment on the execution time}
\label{sec:experiments-time}

We measure the running time $t_o$ spent by the original query and the running time $t_r$ spent by the rewritten query and we compute the corresponding speedup $\dfrac{t_o - t_r}{t_o}$. Time $t_r$ includes the time spent for the selection of a rewriting. 

To evaluate the scalability of our approach, we perform the experiment for different values of the size $s$ of the database and for a fixed value of confidence $c=0.9$ and fixed values of thresholds $T_{corr} =T_{comp}$.

We expect, for the approach to be efficient, that the overhead determined by the procedure of semantic transformation  is by far compensated by the simplifications introduced in the actual execution of the rewritten query.
We expect to observe a positive speedup, as close to 1 as much the optimization proves to be efficient.

Notice that the improvement in the  efficiency of the rewritten query depends by several factors, as the selectivity of the predicates, the relative cost of the eliminated joins -- if any -- w.r.t. the other operations in the query, the size of the set of answers, etc. We might thus perform other experiments to evaluate the amount of saving in execution time as a function not only of the size of the database as before, but also of these aforementioned aspects.

\subsubsection{Experiment on the quality of the answers}
\label{sec:experiments-quality}

The second experiment evaluates the impairment in the quality of the answers retrieved by the rewritten query.

For each query $Q \in \{Q_3, Q_4, Q_7, Q_9, Q_{10}, Q_{15}\}$, we shall execute different tests by varying the minimum value $c$ of confidence of the uncertain integrity constraints that can be possibly used to optimize the query. As mentioned before, we alter the database accordingly.

Let $A$ be the set of answers returned to the original query $Q$ and let $A'$ be the set of answers returned to the rewritten query $Q'$.

We compute the value of recall $R = \dfrac{|A \cap A'|}{|A|}$ and precision $P = \dfrac{|A \cap A'|}{|A'|}$.
We also compute the values of correctness and completeness of the rewritten query $Q'$ and we compare them with the measures $R$ and $P$ to evaluate the strictness of our lower bounds.

Also, for each single answer retrieved by $Q'$, we compute the lower bound on its correctness,  which  is provided to the user. We examine the correspondence between this value and the actual presence of the answer in the set $A$.

As regards the parameters $T_{corr}$ and $T_{comp}$, we can perform the same experiments, to examine their impact on the accuracy of the returned answers, by varying the values of the user's thresholds and fixing values of confidence $c$.

\subsection{Other extensions}
\label{sec:future-work-extensions}

As a possible future work, we may also want to deal with databases with recursive rules and with databases where uncertain information may occur in all of its components: besides mined integrity constraints, even facts  may hold with uncertainty. Also, we might examine the case of facts, rules and integrity constraints with belief interval $[v,w]$ with left and right value not necessarily equal, and not necessarily greater or equal than $0.5$.

\section{Related Work}
\label{sec:relatedWork}

Besides SQO, semantic knowledge provided by integrity constraints is traditionally  exploited in two primary application fields, namely integrity checking and maintenance (see e.g.~\cite{Martinenghi2007} for a complete survey on the classical uses of integrity constraints).

Integrity checking aims at guaranteeing consistency of the data by  detecting violations caused by operations that update the state of the database. In this context, a large amount of works has been traditionally devoted to address the issue of how to identify in an efficient way the operations that can compromise data consistency: the main approaches (e.g., from the precursors~\cite{Nicolas1982,BernsteinBlaustein1982} to~\cite{SadriKowalski1987,Gupta1994,ChristiansenMartinenghi2006}) are based on a proper simplification of the integrity constraints to be checked, and rely on the assumption of total integrity of the database.
As mentioned in the introduction, the hypothesis of the validity of the integrity constraints is, however, in general unrealistic, and violations to the constraints are quite widespread in the common scenario of databases.  Recently~\cite{DeckerMartinenghi2011,Decker2008} have examined the traditional simplification methods for integrity checking, and have proved that for most of them the hypothesis of total integrity can be relaxed by a milder assumption that, basically, prescribes that no new violation is introduced in the database while extant violations can be tolerated. Notably, integrity checking in the presence of inconsistencies provides the benefit of improving the quality of answers returned to users' queries, as it limits, and possibly reduces, the presence of violations in the database, as long as update operations are executed.

As regards the second main field of application of integrity constraints, maintenance, it complements integrity checking in that it does not only check the introduction of inconsistencies caused by updates of the state of the database, but also restores the consistency of the database if an operation that violates the constraints is performed.
Integrity is obtained either by executing a rollback of the operation  or by altering the other components of the database (i.e., by adding or deleting suitable tuples); in this second case the resulting database is called a repair. 
There has been considerable work dealing with the problem of answering queries in databases that violate the integrity constraints, referring to the concept of repair. In particular, \cite{Arenas1999} introduced the notion of Consistent Query Answering (CQA) proposing a method to compute query answers that are consistent, in the sense that they belong to the intersection of the answers to the query on all the possible repairs of the database. The method described in this work is based on a rewriting of the query that uses the set of integrity constraints (which can be possibly violated); other subsequent works address the same problem presenting rewritings for more general patterns of queries~\cite{ArenasBertossiKifer2000,Fuxman2005}, while more general approaches~\cite{Arenas2003,Greco2003} specify database repairs as the the models of a logic program.

Another classical application field of integrity constraints, besides SQO, is cooperative query answering~\cite{Gaasterland1992}, where however the constraints are used only as long as they are valid.

Similarly to the original approaches in the application fields of integrity checking, and also cooperative query answering, the classical works in the area of SQO are based on the assumption of total integrity of the database~\cite{GrantMinker1990,GrantMinker1992}.

There are notably some exceptions to this trend, which instead purposely rely on the presence of inconsistent semantic knowledge to exploit it for query answering, easing the processing of a query. 

An example is again~\cite{Fuxman2005}, which presents a system, called Conquer, which rewrites SQL queries into a form that is able to retrieve consistent answers in relational databases that may violate key constraints. Conquer can annotate the tuples in the database with a Boolean flag that states whether the tuple violates or not a key constraint; the annotations are then used in the phase of query answering to speed up the identification of answers that are inconsistent. 
The exploitation of semantic knowledge is however, to some extent, quite limited.
Differently from our approach, Conquer refers only to the constraints specified in the schema of the relational database (mostly, key constraints) and does not resort to other possible sources of semantic knowledge (as mined association rules) for further semantic optimizations.

The approach that is mostly closest to ours is presented in~\cite{Godfrey2001}. In this work, integrity constraints are distinguished into \emph{informational constraints}, which are required to be never violated, and soft integrity constraints, which can either have no violations in the current state of the database but can be possibly violated in the future  (\emph{absolute soft constraints}, ASCs), or  can already have some violations in the current database state (\emph{statistical soft constraints}, SSCs). SSCs are inferred, e.g., by applying data mining techniques and are associated with statistical information represented by the confidence of the association rule in the database. All the three types of semantic knowledge are exploited for the optimization of a query; however, SSCs  are used to estimate the cardinality of the intermediate results of a query, thus easing the procedure of generating an efficient query plan, but they are  not used for the rewriting of a query (so that the returned answers are guaranteed to be exact). 

Besides~\cite{Godfrey2001}, there are also several works that automatically derive and exploit soft (also called dynamic) constraints for SQO~\cite{Shekhar1993,Siegel1992,Yu1989}, but this semantic knowledge is used only as long as it is true, and then either updated or discarded, so that the information inferred from the database by using it is correct.

In general, to the best of our knowledge, there are no other works that exploit inconsistent semantic knowledge for SQO with the purpose of approximate query answering.

From another point of view, there is a large amount of works related to our approach, which deal in various ways with the representation of inconsistent or conflicting information in databases, but that present some limitations that hinder them from being naturally suitable for approximate SQO.

For instance, \cite{Gatterbauer2009} introduces the model of belief databases, where the tuples of a (relational) database are annotated by users' beliefs, which express their agreement or disagreement on the reliability of the data. Beliefs do not measure uncertainty of knowledge, but provide a qualitative evaluation of the presence of conflicts in the information contained in the database: thus, they are taken into account for query answering, but they are not exploited to optimize a query or to provide approximate results.

Another classical model proposed to manage uncertain knowledge is based on probabilistic databases~\cite{Dalvi2007}, also extended to inconsistent probabilistic databases~\cite{Lian2010}; among the models of uncertainty we can also mention probabilistic logic programming~\cite{Lukasiewicz2001,Ng1993,Deraedt2008}: however,  differently from our analysis of inconsistent information in the theory of BLP, all of them typically rely on some loose hypotheses on the independence of sources of knowledge that disregard possible correlations in the data.

From another point of view, there is also some interesting work on the automatic detection of violations to integrity constraints by applying data mining techniques, but with purposes different from query optimization: e.g.,~\cite{Ceri2007} introduces the notion of pseudoconstraints (which are exploited for the identification of rare events in databases, as the presence of outliers) and~\cite{Mazuran2009}, which mines violations to integrity constraints that are not valid to update and relax them.

Finally, as regards the traditional approaches for approximate query answering (as~\cite{Chaudhuri2001,Babcock2003,Spiegel2009}), they mostly provide fast, not exact, answers relying on the computation of suitable synopses of the data (see also, e.g.,~\cite{Ilyas2008}, for a comparative survey of works on approximate top-k query processing): as a future work we plan to perform our campaign of experiments and compare the results of our approach to the current state-of-the-art approaches.

\section{Conclusions}
\label{sec:conclusions}

This report described an approach to generate approximate answers to queries to a possibly inconsistent database by exploiting not necessarily valid semantic knowledge.
Our approach relies on the classical technique of SQO, which uses integrity constraints to rewrite a query into a more efficient form, and we adapted the traditional approach to handle also inconsistent integrity constraints. Also, we applied concepts from the theory of Belief Logic Programming to deal with the presence of possible correlation in the semantic knowledge used to optimize a query.

The approach for approximate SQO is first presented for the simplified case of a database with only extensional predicates, and then we extended the presentation to the more general case of a database with a non-empty IDB, where the rules, integrity constraints or queries cannot contain negated intensional predicates. We presented some hints on how to deal with the presence of possibly negated intensional predicates.

We mentioned some future work in Section~\ref{sec:future-work}: namely, we plan to perform a set of experiments to validate our approach, and we described in that section the steps of the experimental evaluation that we intend to accomplish; also, we plan to generalize the procedure to handle to a greater extent the presence of uncertainty in the database (dealing also with the presence of uncertain facts or with less reliable rules or integrity constraints).

\bibliographystyle{plain}
\bibliography{minor}

\end{document}